\documentclass{elsart}
\usepackage{natbib}
\usepackage{graphicx}
\usepackage{amssymb}
\usepackage{amsmath}

\numberwithin{equation}{section}

\newcommand{\bv}{\boldsymbol{b}}

\newcommand{\fv}{\boldsymbol{f}}
\newcommand{\nv}{\boldsymbol{n}}
\newcommand{\rv}{\boldsymbol{r}}
\newcommand{\uv}{\boldsymbol{u}}
\newcommand{\Jv}{\boldsymbol{J}}

\newcommand{\Lamv}{\boldsymbol{\Lambda}}
\newcommand{\lamv}{\boldsymbol{\lambda}}

\newcommand{\Fc}{\mathcal{F}}
\newcommand{\Lc}{\mathcal{L}}
\newcommand{\Uc}{\mathcal{U}}
\newcommand{\Hc}{\mathcal{H}}
\newcommand{\Ac}{\mathcal{A}}
\newcommand{\Bc}{\mathcal{B}}
\newcommand{\Pc}{\mathcal{P}}
\newcommand{\Ic}{\mathcal{I}}

\newcommand{\eps}{\varepsilon}

\newcommand{\Gr}{\mathrm{Gr}}
\newcommand{\Pra}{\mathrm{Pr}}
\newcommand{\Rey}{\mathrm{Re}}
\newcommand{\Rm}{\mathrm{Rm}}

\newcommand{\Rot}{\nabla\times}
\newcommand{\Div}{\nabla\cdot}

\newcommand{\uvKol}{\boldsymbol{u}_{\mathrm{Kol}}}
\newcommand{\sKol}{s_{\mathrm{Kol}}}
\newcommand{\OmegaKol}{\Omega_{\mathrm{Kol}}}

\newcommand{\fvt}{\widetilde{\fv}}
\newcommand{\Phiv}{\boldsymbol{\Phi}}

\newcommand{\dx}{\, dV}
\newcommand{\Fv}{\boldsymbol{F}}

\newcommand{\Tbb}{\mathbb{T}}

\def\rf#1{(\ref{#1})}

\begin{document}

\begin{frontmatter}

\title{Dissipative structures in a \\ nonlinear dynamo}

\author[label1]%
{Andrew D. Gilbert,}
\author[label2]%
{Yannick Ponty}
\author[label3]%
{and Vladislav Zheligovsky}

\address[label1]{Mathematics Research Institute,
University of Exeter, U.K.}

\address[label2]{Observatoire de la C\^ote d'Azur, Nice, France.}

\address[label3]{International Institute of Earthquake Prediction Theory\\
     and Mathematical Geophysics, Moscow, Russia.}

\begin{abstract}

This paper considers magnetic field generation by a fluid flow in a system referred to as the Archontis dynamo: a steady nonlinear magnetohydrodynamic (MHD) state is driven by a prescribed body force. The field and flow become almost equal and dissipation is concentrated in cigar-like structures centred on straight-line separatrices. Numerical scaling laws for energy and dissipation are given that extend previous calculations to smaller diffusivities. The symmetries of the dynamo are set out, together with their implications for the structure of field and flow along the separatrices. The scaling of the cigar-like dissipative regions, as the square root of the diffusivities, is explained by approximations near the separatrices. Rigorous results on the existence and smoothness of solutions to the steady, forced MHD equations are given.

\end{abstract}

\end{frontmatter}

\section{Introduction}\label{secintro}

Much is known about fast dynamo action: the rapid growth of magnetic fields at high magnetic Reynolds number in fluid flows with chaotic streamlines, but the mechanisms for the dynamical saturation of such fields remain poorly understood. In many cases when the growing field equilibrates by modifying the fluid motion, the effect is to switch off the chaotic stretching in the flow, as measured for example by a reduction in the finite-time Liapunov exponents \cite[e.g.,][]{CaHuKi96,ZiPoPo98}. What is left is a fluid threaded by a magnetic field which resists stretching and so suppresses overturning fluid motions, but supports elastic wave-like motions, essentially Alfv\'en waves with coupled field and flow \cite[e.g.,][]{CoHuPr10}. The final state of many simulations shows apparently chaotic behaviour in space and time, suggestive of an attractor of moderate or high dimension, although because of the three-dimensionality of MHD systems little can be done to explore its properties, for example the fractal dimension or spectrum of Liapunov exponents.

Although this appears to be the outcome of many simulations, as far as they can be run, there are some intriguing examples where a further phase of evolution takes place: the magnetic field and flow align, depleting the nonlinear terms, and both fields evolve to a steady (or very slowly evolving) state. The key point is that in unforced, ideal magnetohydrodynamics (see equations (\ref{eqNS0}--\ref{eqdivfree0}) below with $\nu=\eta=0$ and $\fv=0$) any state with $\uv = \pm \bv$ is an exact steady solution. The remarkable fact that simulations of forced, non-ideal MHD turbulence could evolve to something very close to such a state was first observed by \cite{Ar00} in his thesis, and published in \cite{DoAr04} (hence referred to as DA), and \cite{ArDoNo07}. These simulations use a compressible code with a Kolmogorov forcing function, (\ref{eqSvdef}) below, first used as the form of a flow for simulations of fast, kinematic dynamo action by \cite{GaPr92}. Subsequently \cite{CaGa06a} undertook incompressible simulations of the same system as Archontis, and pushed up the fluid and magnetic Reynolds numbers; our work is linked closely to this paper, which we refer to as CG in what follows.

What these authors found was that, starting with a forced fluid flow and a seed magnetic field, the growing magnetic field initially equilibrates in rough equipartition with the velocity field, in a messy, chaotic time-dependent state. However during this state, there is a slow but persistent exponential growth in the average alignment of the $\uv$ and $\bv$ vectors, as measured by the cross-helicity. This process of alignment continues until there takes place a sudden increase in the fluid and magnetic energies, and both fields tend to a steady state of almost perfect alignment, discrepancies being controlled by the weak dissipation and the forcing. In fact since any solution $\uv=\pm\bv$ is a neutrally stable solution of the ideal problem \citep{FrVi95}, the solution that is selected must depend delicately on balances involving these subdominant diffusive and forcing effects. We note that some alignment of field and flow has been noted in many other MHD flows, for example see \cite{DoMaVe80}, \cite{PoMeFr86}, \cite{MaCaBo06} and references therein, but of a less dramatic nature.

This observation of dynamo saturation in a steady state with such a high degree of alignment was a new phenomenon: CG refer to the saturated state as the `Archontis dynamo', though we prefer the term `Archontis saturation mechanism'. CG observed this aligned  state as a solution branch over a wide range of magnetic and fluid Reynolds numbers (taking the magnetic Prandtl number to be unity in much of their work). Further developments include the development of bursts of rapid time dependence after some time in the steady state, in the study \cite{ArDoNo07}. However this appears only to occur in the compressible case, as it has not been seen by CG nor in our simulations; we will therefore not discuss this further. \cite{CaGa06b} also find slow time-dependent evolution of the saturated state for the Kolmogorov forcing with magnetic Prandtl number $\Pra = \nu/\eta$ not equal to unity, and for more general spatially periodic steady forcings. In all cases though, the field and flow settle into a state of very close alignment, even if they then evolve on a slow time scale.

The focus of the present paper is to understand more about the structure of the steady saturated state for the Kolmogorov forcing and unit magnetic Prandtl number $\Pra$, with a particular focus on the regions where dissipation occurs and on rigorous results on existence and smoothness. DA and CG find a complex geometrical picture for the field and flow and identify these regions of high dissipation: they are localised along straight-line separatrices that join a family of stagnation points; similar structures are found in the
1:1:1 ABC flow \citep{DoFrGrHeMeSo86}. These are found to have a width scaling as $\sqrt{\eps}$ where $\eps$ is a dimensionless measure of the diffusivity, and one of our aims is to understand this power law.

We set up the governing equations in \S\ref{secgov} and extend the solution branch to yet smaller values of the diffusivity $\eps$ by means of large scale simulations in \S\ref{secnum}. In \S\ref{secsym} we then classify the symmetries of the Kolmogorov forcing, which are preserved by the nonlinear, saturated field and flow. These symmetries are the reason for the presence of the non-generic straight line separatrices that join stagnation points in the flow and field, and they constrain the local flow: it is in these regions that dissipation is strongest. We plot the local structure of fields along the separatrix from $(0,0,0)$ to $(\pi,\pi,\pi)$ in \S\ref{secflowfield}. We determine the effects of diffusion by setting up PDEs for the advection of field as it enters the dissipative regions in \S\ref{secPDE} and use these to justify the order $\sqrt{\eps}$ scaling for the cigar widths found in CG.
We then proceed with a formal mathematical investigation of the existence of steady-state solutions to the MHD problem at hand and bounds for them in various function spaces in \S\S \ref{maths}--\ref{smooth}. The reader should note that these sections use functional analysis and so have a different flavour from the earlier ones.
%
% (the reader should be forewarned that the material of these sections employs more advanced mathematical tools of functional analysis than those used in other, more physics-oriented sections).
Finally \S\ref{secdisc} offers concluding discussion.

\section{Governing equations}\label{secgov}

We begin with the dimensional equations for incompressible MHD, in the form
\begin{align}
\partial_t \uv + \uv \cdot\nabla\uv & = \bv \cdot\nabla\bv -\nabla p + \nu \nabla^2 \uv + \fv, \label{eqNS0}\\
\partial_t \bv + \uv\cdot\nabla\bv &=  \bv \cdot\nabla\uv + \eta \nabla^2\bv ,
\label{eqinduc0}\\
\nabla\cdot \uv & = \nabla\cdot\bv = 0 ,
\label{eqdivfree0}
\end{align}
where $\nu$ and $\eta$ are the kinematic viscosity and magnetic diffusivity.
We take $\fv$ to be a steady force of magnitude $\Fc$ acting on a length scale $\Lc$. We will consider the \emph{Kolmogorov forcing} $\fv = \Fc\fv^*(\rv/\Lc)$, whose dimensionless form is given by
\begin{equation}
\fv^*(\rv) = (\sin z, \sin x, \sin y) .
\label{eqSvdef}
\end{equation}

In non-dimensionalising we have only the parameters $\{\Lc,\Fc,\nu,\eta\}$, together with the form (\ref{eqSvdef}) of the forcing function. From these we can define a magnetic Prandtl number and a Grashof number as in similar forced flow problems \citep[see, e.g.,][]{ChKeGi01} by
\begin{equation}
\Pra = \nu/\eta, \quad
\Gr = \Fc\Lc^3/\nu^2\equiv\eps^{-1}   .
\end{equation}
We have as diagnostics the Reynolds number and magnetic Reynolds number given by
\begin{equation}
\Rey = \Lc\lVert\uv\rVert/\nu, \quad \Rm = \Lc\lVert\uv\rVert/\eta ,
\end{equation}
where $\lVert\uv\rVert$ is a measure of the fluid velocity at a given time, for example the $L^2$ norm, taken as the root-mean-square value, averaged over the periodicity box. We rescale as
\begin{equation}
\uv = \Uc \uv^*, \quad
\bv = \Uc \bv^*, \quad
t = (\Lc/\Uc) t^*, \quad
\rv = \Lc\rv^*, \quad
\fv = \Fc\fv^*, \quad
p = \Uc^2 p^*,
\end{equation}
with the choice of velocity scale
\begin{equation}
\Uc = \Fc\Lc^2/\nu .
\end{equation}
This yields the non-dimensional formulation, dropping the stars, as
\begin{align}
\partial_t \uv +\uv \cdot\nabla\uv &= \bv \cdot\nabla\bv -\nabla p + \eps \nabla^2 \uv + \eps\fv,
\label{eqNS}\\
\partial_t \bv+\uv\cdot\nabla\bv  &=  \bv \cdot\nabla\uv + \eps \Pra^{-1}\nabla^2\bv ,
\label{eqinduc}\\
\nabla\cdot \uv & = \nabla\cdot\bv = 0 ,
\label{eqdivfree}
\end{align}
with $\fv$ given in (\ref{eqSvdef}) and the only parameters specified are $\{\eps,\Pra\}$. The corresponding Reynolds and magnetic Reynolds numbers are
%then time-dependent diagnostics, to be measured as
%
\begin{equation}
\Rey = \eps^{-1}\lVert\uv\rVert, \quad \Rm = \eps^{-1}\Pra\lVert\uv\rVert.
\end{equation}
We refer to $\eps^{-1}$ as the Grashof number $\Gr$ and will be interested in the inviscid limit  $\eps\to0$. The Reynolds number and magnetic Reynolds number are diagnostics depending on the flow regime realised.%
\footnote{Our formulation is equivalent to DA/CG, but our terminology is a little different. For example CG use the parameters
$\nu_{\text{CG}}\equiv \eps$ and $\eta_{\text{CG}}\equiv\eps\, \Pra^{-1}$,
which they refer to as the inverse Reynolds and magnetic Reynolds numbers respectively.}
Indeed, they change greatly during the saturation process, when the fields align and $\lVert\uv\rVert$, $\lVert\bv\rVert$ increase significantly.
As in CG, the governing equations may be written in a more symmetrical form in terms of Elsasser variables
\begin{equation}
\Lamv_\pm  = \uv\pm\bv,
\label{eqElsdef}
\end{equation}
which gives, for $\Pra = 1$,
\begin{align}
\partial_t \Lamv_+ + \Lamv_- \cdot\nabla\Lamv_+  &=  -\nabla p + \eps \nabla^2 \Lamv_+ + \eps\fv, \label{eqElsplus}\\
\partial_t \Lamv_- + \Lamv_+ \cdot\nabla\Lamv_-  &=  -\nabla p + \eps \nabla^2 \Lamv_- + \eps\fv, \label{eqElsminus}\\
\nabla\cdot \Lamv_+ =\nabla\cdot\Lamv_- & = 0 .\label{eqElsdiv}
\end{align}

\section{Numerical results}\label{secnum}

We undertook a number of runs to investigate the structure of the steady, equilibrated Archontis dynamo for $\Pra=1$ and values of $\eps$ down to $10^{-4}$ in the $(2\pi)^3$ periodic domain $\Tbb^3$. The steady solutions were found by following the solution branch: that is taking the output from a run with a given value of $\eps$ and using it as the initial condition for a run with a reduced value of $\eps$. This establishes the Archontis dynamo as a robust local attractor, in the range of $\eps$ used, in agreement with DA and CG. Whether it is a global attractor over some or all  sufficiently small values of $\eps$ remains unknown, and extremely difficult to address in view of the long transients that may occur. Our runs were undertaken with a pseudo-spectral code using $N^3$ modes with $N= 128$ for $\eps = 0.02$ and $0.01$, $N=256$ for $\eps=10^{-3}$, and $N=512$ for $\eps=10^{-4}$. There were other, less well resolved runs with $N=128$ for $\eps=10^{-3}$ and $N=256$ for $\eps=10^{-4}$,  which we refer to below as our `testing simulations'. For comparison, CG go down to $\eps=1.25\times10^{-3}$ in their study, with resolution $128^3$. Our results thus extend theirs by a little over a decade, and in this section we present measures of the magnetic field and flow in the equilibrated state.

\begin{figure}
\includegraphics[scale=0.6]{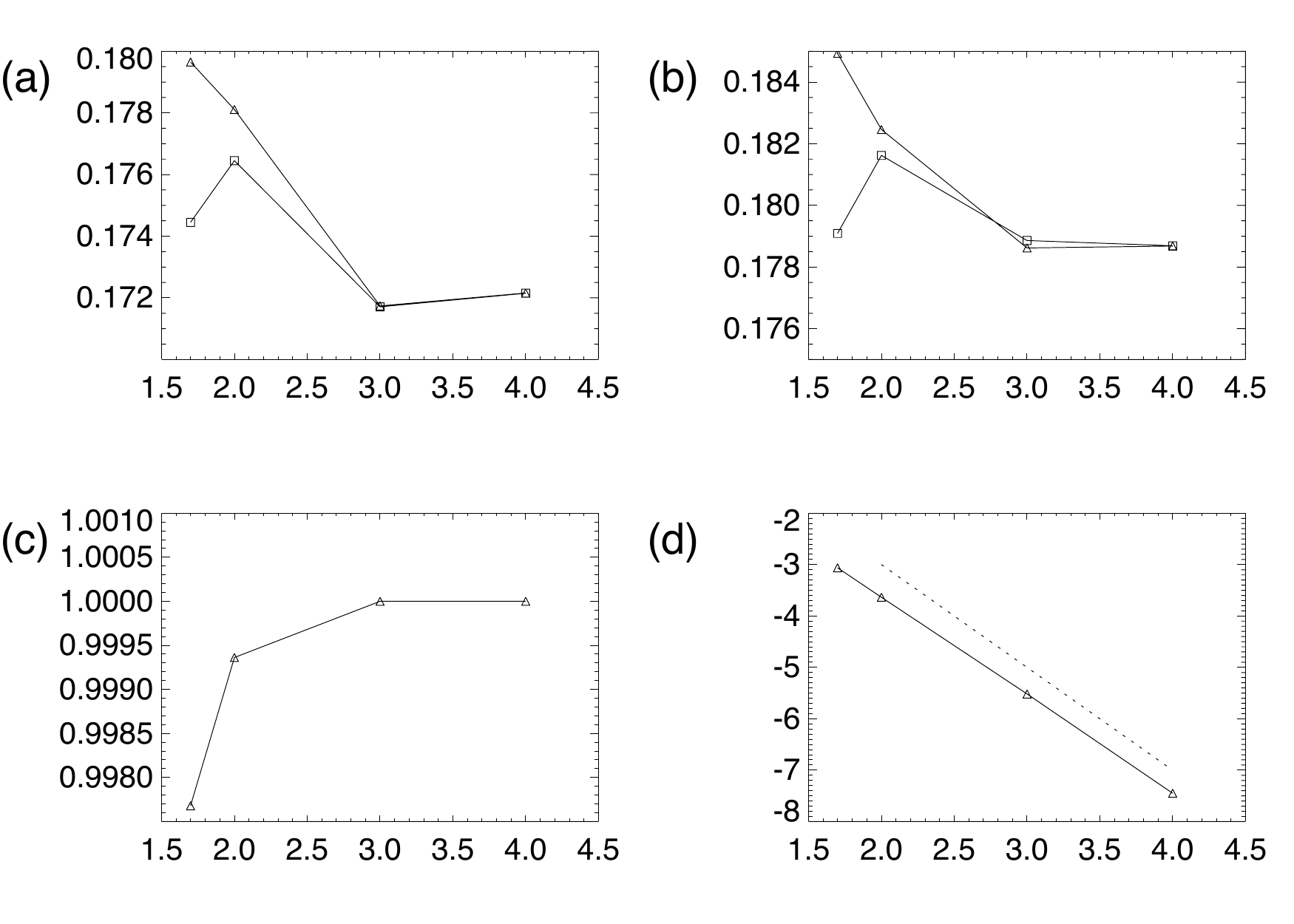}
%\vspace{5cm}
\caption{Numerical results plotted against $\log_{10} \eps^{-1}$. Plotted are
(a) kinetic energy $E_K$ (triangles) and magnetic energy $E_M$ (squares),
(b) enstrophy $\Omega_K$ (triangles) and squared current $\Omega_M$ (squares),
(c) normalised cross helicity $H_X/(2E_K E_M) $,
(d) energy $\log_{10} E_-$ of $\Lamv_-$ (dotted line gives $\eps^2$ dependence). }
\label{figplotEmEv}
\end{figure}

\begin{table}
\begin{tabular}{|c|c||c|c|c|c|c|c|c|}
\hline
 $\eps$ & $N$ & $E_K$ & $E_M$ & $\Omega_K$ & $\Omega_M$ & $H_X$ & $E_-$ \\
\hline
0.02     & 64   & 0.1797 & 0.1745 & 0.1849 & 0.1791 & 0.3532 & $8.685\times10^{-4}$ \\

0.01     & 64   & 0.1781 & 0.1765 & 0.1825 & 0.1816 & 0.3543 & $2.313\times10^{-4}$\\

0.001   & 256 & 0.1717 & 0.1717 & 0.1786 & 0.1789 & 0.3435 &  $3.04\times10^{-6}$  \\

0.0001 & 512 &  0.1722 & 0.1722 & 0.1787 & 0.1787 & 0.3443 & $3.55\times 10^{-8}$ \\
 \hline
\end{tabular}
 \medskip
\caption{Numerical results. }
\end{table}
%
%%
%\begin{table}
%\begin{tabular}{|c|c|c|c|c|c|c|c|c|}
%\hline
% $\eps$ & $N$ & $E_K$ & $E_M$ & $\Omega_K$ & $\Omega_M$ & $H_X$ & $E_-$ \\
%\hline
%0.02     & 64   & 0.17965 & 0.17445 & 0.18493 & 0.17909 & 0.35324 & $8.6849\times10^{-4}$ \\
%
%0.01     & 64   & 0.17811 & 0.17645 & 0.18246 & 0.18162 & 0.35433 & $2.3131\times10^{-4}$\\
%
%0.001   & 128 & 0.17132 & 0.17129 & 0.17800 & 0.17825 & 0.34261 & $3.0458\times10^{-6}$ \\
%
%0.001   & 256 & 0.17174 & 0.17171 & 0.17862 & 0.17886 & 0.34345 &  $3.04\times10^{-6}$  \\
%
%0.0001 & 256 & 0.17216 & 0.17216 & 0.18204 & 0.18207 & 0.34432 & $3.66\times 10^{-8}$ \\
%
%0.0001 & 512 &  0.17216 & 0.17215 & 0.17868 & 0.17869 & 0.34431 & $3.55\times 10^{-8}$ \\
% \hline
%\end{tabular}
% \medskip
%\caption{Numerical results: full results not for publication! These are 5SF but I do not trust the last one so the proper results in table 1 are rounded. . ? means YP to supply. }
%\end{table}
%%

Numerical values are given in table 1 and plotted in figure \ref{figplotEmEv}. Panel \ref{figplotEmEv}(a) shows the kinetic and magnetic energies in the equilibrated state, given by
\begin{equation}
E_K = \int_{\Tbb^3} \tfrac{1}{2} |\uv|^2\, dV,
% \equiv \tfrac{1}{2} (2\pi)^3 \lVert \uv\rVert^2,
\quad
E_M = \int_{\Tbb^3} \tfrac{1}{2} |\bv|^2\, dV .
%\equiv \tfrac{1}{2} (2\pi)^3 \lVert \bv\rVert^2 ,
\end{equation}
These show an initial decrease with $\eps$ (as in CG) but then a slight increase from $\eps = 10^{-3}$ to $\eps = 10^{-4}$: this is quite small bearing in mind the scale on the vertical axis, but appears to be real as it is borne out in our test simulations. In all these runs $E_K>E_M$ though this is not apparent from the numbers in table 1 nor in panel \ref{figplotEmEv}(a). Panel \ref{figplotEmEv}(b) shows the enstrophy and integrated squared current, defined by
\begin{equation}
\Omega_K = \int_{\Tbb^3} \tfrac{1}{2} |\nabla\times\uv|^2\, dV ,
\quad
\Omega_M = \int_{\Tbb^3} \tfrac{1}{2} |\nabla\times\bv|^2\, dV .
\end{equation}
The total dissipation is given by $2\eps\Omega_K+2\eps\Omega_M$ and this tends to zero as $O(\eps)$, as does the input of mechanical energy. Panel  \ref{figplotEmEv}(c) shows the cross helicity
\begin{equation}
 H_X =  \int_{\Tbb^3} \uv\cdot\bv\, dV
\end{equation}
in normalised form, which rapidly tends to its theoretical upper bound of unity, within the accuracy of our simulations, indicating the strong alignment of field as $\eps\to0$. Finally panel \ref{figplotEmEv}(d) shows the energy in the $\Lamv_-$ Elsasser variable, where
\begin{equation}
E_{\pm} = \int_V \tfrac{1}{2} |\Lamv_\pm|^2\, dV \equiv E_K + E_M \pm H_X.
\end{equation}
This shows a rapid decrease to zero as $\eps\to0$ consistent with the scaling $E_- \propto \eps^2$ (dotted line) in agreement with the discussion in CG and below.%
\footnote{The (downwards) slope of a line fitting the data points is close to $1.9$: the reason for the discrepancy is unclear: it could be numerical, or the $\eps^2$ power law may only be achieved as $\eps\to0$ which is possible as there is some downwards curvature present in the data points.}

\begin{figure}
\includegraphics[scale=0.6]{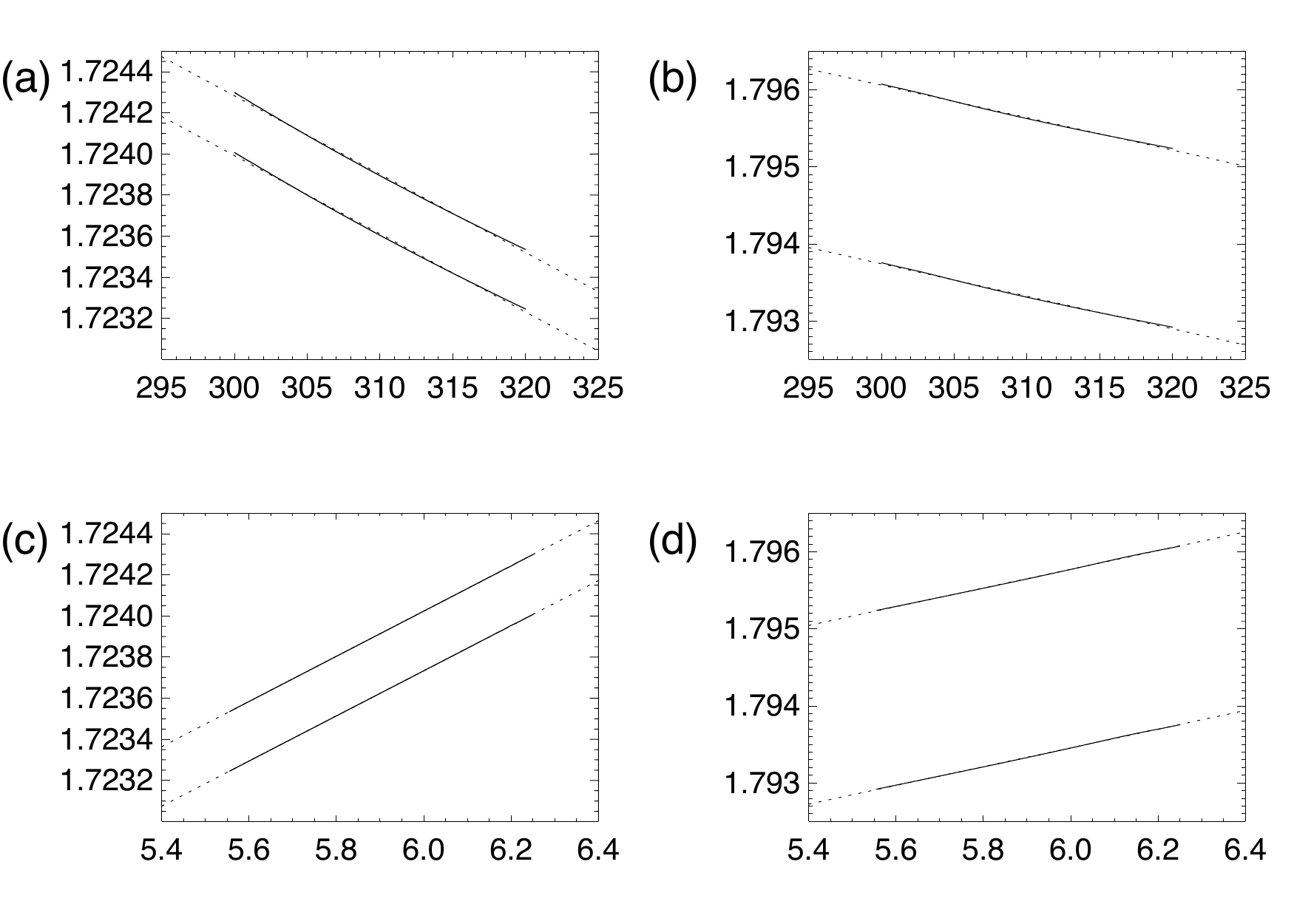}
%\vspace{5cm}
\caption{Numerical results for the case $\eps=10^{-3}$ with $N=256$.
Plotted are
(a) $10E_K$ (upper) and $10E_M$ (lower),
(b) $10\Omega_K$ (lower) and $10\Omega_M$ (upper), against $t/10$.
(c,d) are the same but plotted against $10^4/(t-1400)$.
In each panel dotted lines are linear fits.  }
\label{figfitEmEv}
\end{figure}

In panel  \ref{figplotEmEv}(b) it is notable that the two curves, for enstrophy and total current squared, cross between $\eps=0.01$ and $0.001$. The enstrophy $\Omega_K$ is a little smaller than $\Omega_M$ for $\eps=10^{-3}$ and in fact is also for $
10^{-4}$ and in our test simulations, making us confident that this is a real effect. This opens up the question of how we measure these quantities, since the rate of evolution of the state becomes extremely slow for small $\eps$. Figure \ref{figfitEmEv}(a,b) shows $E_K$, $E_M$, $\Omega_K$ and $\Omega_M$ as functions of time for the case $\eps=10^{-3}$ and $N=256$: comparison with linear fits (dotted) shows clear curvature, as expected, but also highlights the slow evolution. This suggests neutral stability of the final state, and an expansion for any quantity in the form
\begin{equation}
A = A_0 + A_1 t^{-1} + A_2 t^{-2} + \cdots .
\end{equation}
Although asymptotically the origin of time does not matter, we found it helpful to choose an origin of time $t_0$ (once per run) so as to obtain the best linear fit for quantities in the form
\begin{equation}
A \simeq A_0 + A_1 (t-t_0)^{-1}
\end{equation}
We then use an estimate of the limiting value as $A_0$; for example see figure \ref{figfitEmEv} (c,d). This was done for all the results in table 1.

\begin{figure}
(a)\includegraphics[scale=0.39]{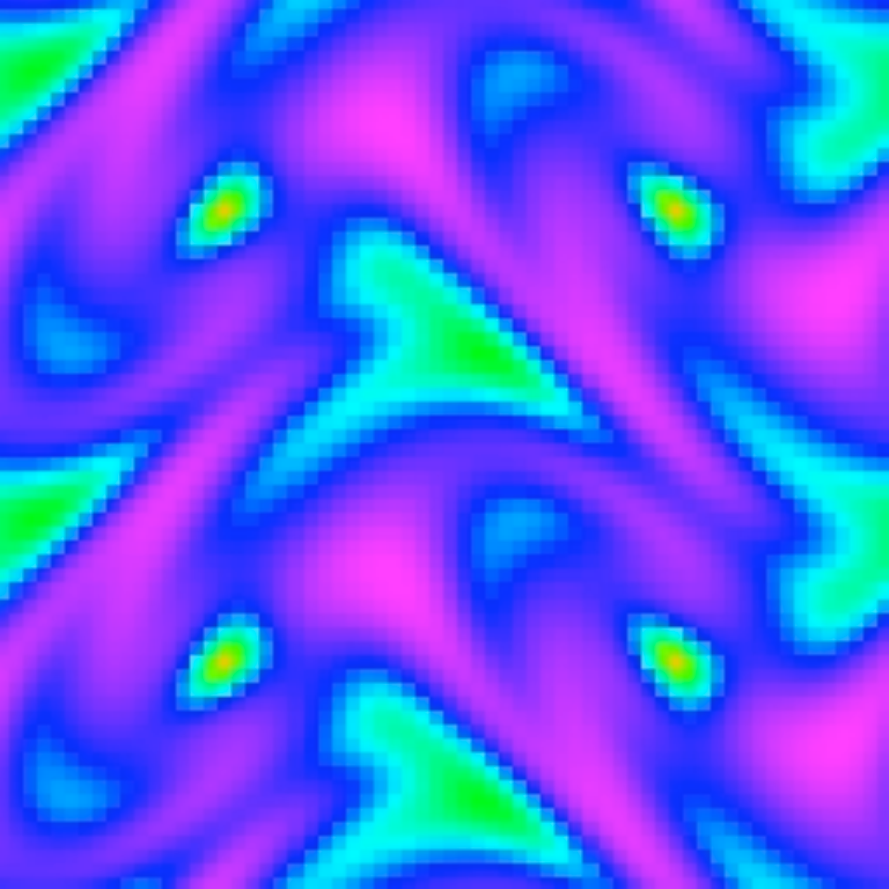}
(b)\includegraphics[scale=0.39]{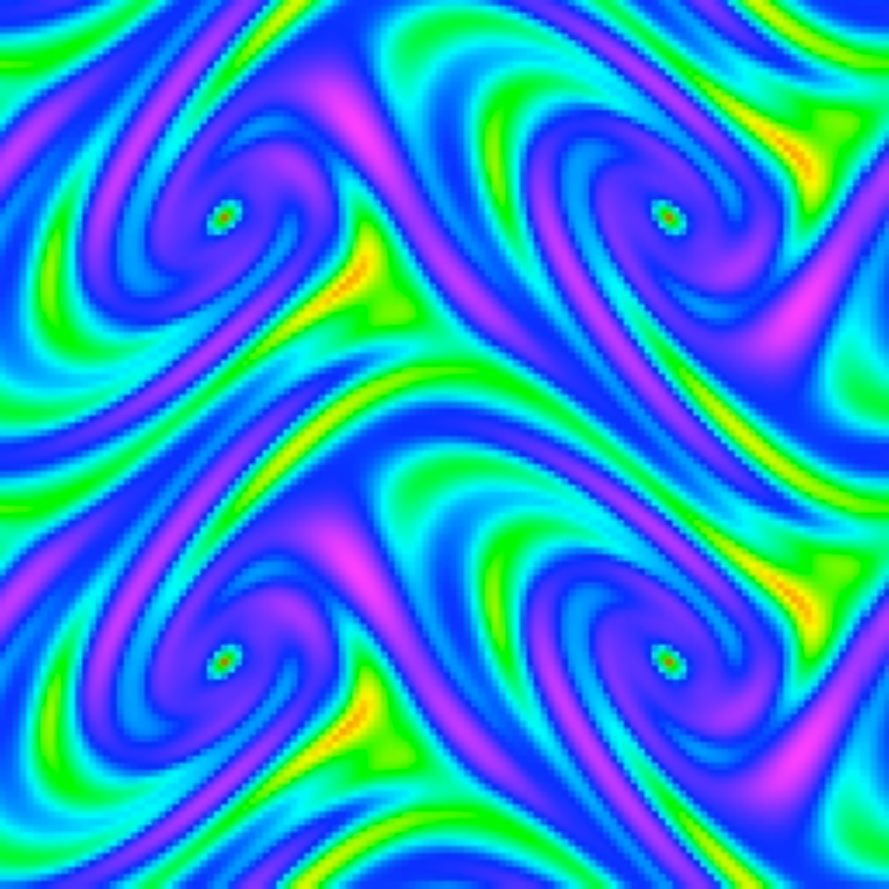}
(c)\includegraphics[scale=0.39]{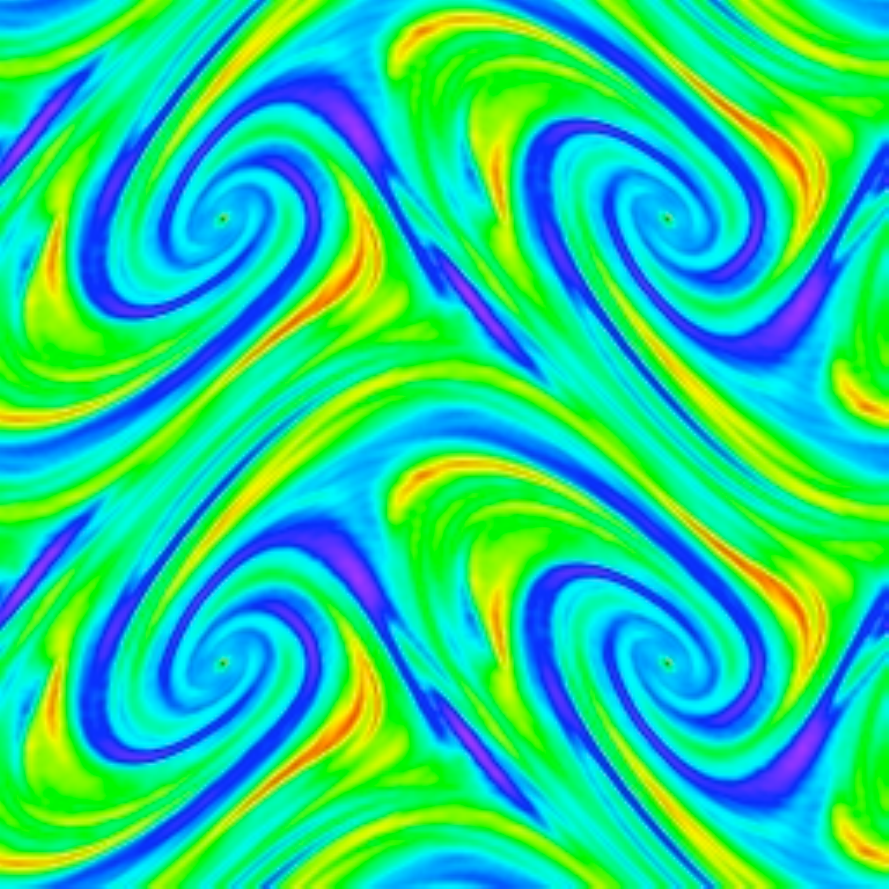}
\includegraphics[scale=0.214]{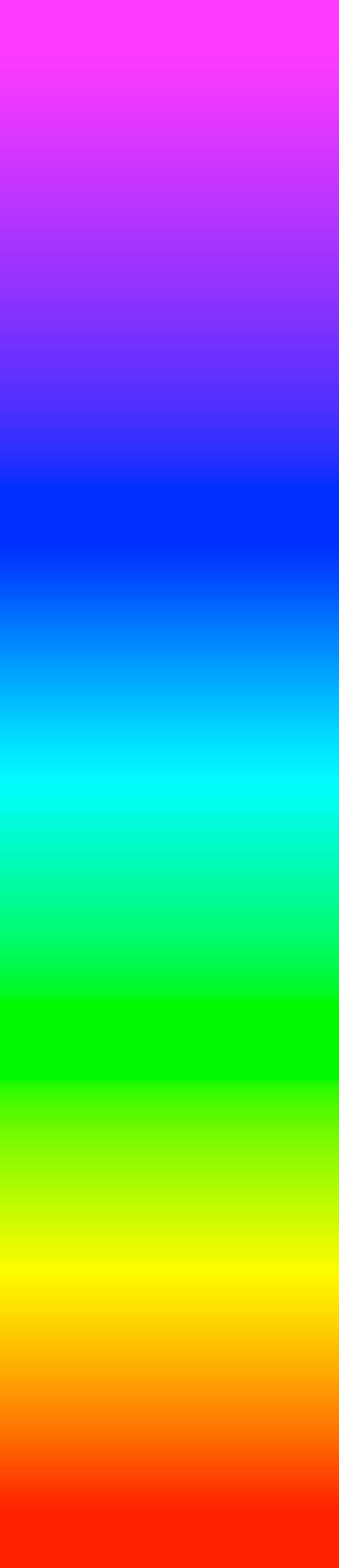}
%\vspace{5cm}
\caption{Cross sections showing $|\Lamv_-|$ in the $(x,y)$-plane for $z= \pi/2$. In (a) $\eps=10^{-2}$, (b) $10^{-3}$ and (c) $10^{-4}$ and the colour scale shown runs from zero (bottom) to (a) 0.031, (b) 0.0041 and (c) 0.00058 (top). }
\label{figvapor}
\end{figure}

One of the aims of this paper is to focus on dissipative regions in the system: these occur along a series of straight line separatrices (DA/CG) and in figure \ref{figvapor}, we show colour plots of $|\Lamv_-|$ for a range of diffusivities. Clearly seen in each case, but especially in (c) at the smallest $\eps$, are cross sections of spiralling field, centred on the separatrices, where small scales are generated with consequently enhanced diffusion.

%--------------------------------------------------------------------------------------------

\section{Symmetries}\label{secsym}

We have seen that the dissipation tends to concentrate in cigar shaped regions, with one extending from $(0,0,0)$ to the stagnation points at $\pm(\pi,\pi,\pi)$. The reason these straight line separatrices are robust structures is linked to the symmetries of the forcing (\ref{eqSvdef}) and also applies to the kinematic dynamo study by \cite{GaPr92} of the Kolmogorov flow
\begin{equation}
\uvKol(\rv) = (\sin z, \sin x, \sin y) .
\label{eqKol}
\end{equation}
These symmetries turn out to be preserved by the solution  $(\uv,\bv)$ in the nonlinear regime: there is no symmetry breaking. The forcing (\ref{eqSvdef}) is $2\pi$-periodic in each coordinate: we only
consider symmetries up to this periodicity (and that do not reverse time). Note first that any map $T$ maps a vector field $\uv$ according to
\begin{equation}
(T\uv)(\rv)= \Jv_T\cdot\uv(T^{-1}\rv) , \quad
\Jv_T=\partial \rv/\partial T^{-1}\rv .
\end{equation}
It is then easily checked that the forcing $\fv$ in (\ref{eqSvdef}) is preserved by the following
12 orientation-preserving symmetries, with $\det \Jv = 1$, which form the group $A_4$ of even permutations of 4 objects, or the symmetry group of the tetrahedron,
\begin{align}
  i(\rv)&=(x,y,z),
& a^2(\rv)&=(-x, \pi-y, z+\pi) , \notag\\
 b^2(\rv)&=(x+\pi, -y, \pi-z) ,
& c^2(\rv)&=(\pi-x, y+\pi, -z) ,  \notag\\
  d(\rv)&=(z,x,y),
& d^2(\rv)&=(y,z,x), \label{eqsym}\\
  e(\rv)&=(-z, \pi-x, y+\pi),
& e^2(\rv)&=(\pi-y, z+\pi, -x),  \notag\\
  f(\rv)&=(z+\pi, -x, \pi-y),
& f^2(\rv)&=(-y, \pi-z, x+\pi),  \notag\\
  g(\rv)&=(\pi-z, x+\pi, -y),
& g^2(\rv)&=(y+\pi, -z, \pi-x) . \notag
\end{align}
These also form a subgroup of the group of 24 symmetries of the $1$:$1$:$1$ ABC flow \citep{ArKo83,DoFrGrHeMeSo86}, and the above follows the notation in \cite{Gi92}. The symmetries all commute with the inversion symmetry $j(\rv)=(-x,-y,-z)$ and so the full symmetry group of the forcing $\fv$ is the direct product $A_4\times \mathbb{Z}_2$.

\section{Flow and field on the separatrices}\label{secflowfield}

The above symmetries constrain the behaviour of the magnetic field and flow on the separatrices. Take, for definiteness, the separatrix joining $(0,0,0)$ to $(\pi,\pi,\pi)$ and call this the `main separatrix' for brevity. Because of the symmetries $d$ and $d^2$ in (\ref{eqsym}) there is a three-fold rotational symmetry about this separatrix, as seen in DA/CG,  and any vector field on the separatrix can only point along the separatrix. We may introduce rotated Cartesian coordinates via
\begin{equation}
\begin{pmatrix}
\mu \\ \chi \\ \zeta
\end{pmatrix}
=
\begin{pmatrix}
1/\sqrt{2} &-1/\sqrt{2} & 0 \\
1/\sqrt{6} & 1/\sqrt{6} &-2/\sqrt{6} \\
1/\sqrt{3} & 1/\sqrt{3} & 1/\sqrt{3} \\
\end{pmatrix}
\begin{pmatrix}
x \\ y \\ z
\end{pmatrix} ,
\label{eqcoorddef}
\end{equation}
with $\zeta$ along the separatrix. From there we may further define cylindrical polar coordinates $(\rho,\theta,\zeta)$, whose axis is $\zeta$ along the separatrix with $\mu = \rho\cos\theta$ and $\chi = \rho\sin\theta$.

Our aim now is to investigate more of the behaviour of the flow near to the separatrix, in the saturated regime. However to fix ideas and establish a benchmark, we consider first the Kolmogorov flow $\uvKol$ in (\ref{eqKol}). For this flow it can be shown that on the main separatrix motion is governed by
\begin{equation}
\dot\zeta = \sqrt{3} \sin (\zeta/\sqrt{3}), \quad \mu=\nu= 0,
\end{equation}
with solution
\begin{equation}
\zeta = \sqrt{3} (\pi - \cos^{-1}\tanh t) .
\end{equation}
Here $\zeta$ tends to zero as $t\to-\infty$ and to $\sqrt{3}\pi$ as $t\to\infty$.
%, the midpoint being crossed at $t=0$.
Near to the separatrix, the radial coordinate $\rho\ll1$ and the flow field may be expanded in powers of $\rho$. In view of the three-fold rotational symmetry, the flow $\uv$ is axisymmetric about the main separatrix $\rho=0$ at leading order and streamlines are given by
\begin{equation}
\dot \rho = - s'(\zeta) \rho + O(\rho^2), \quad
\dot\theta = \Omega(\zeta) + O(\rho), \quad
\dot \zeta = 2s(\zeta) + O(\rho^2),
\label{eqsepmotion}
\end{equation}
with
\begin{equation}
2\sKol(\zeta)  = \sqrt{3}\sin(\zeta/\sqrt{3}) , \quad
2\OmegaKol(\zeta) = \sqrt{3}\cos(\zeta/\sqrt{3}) .
\label{eqsepmotionfuncs}
\end{equation}
Trajectories spiral in for $\zeta\simeq 0$ and spiral out for $\zeta\simeq \sqrt{3}\pi$. On the separatrix itself $\uv = (0,0,2s(\zeta))$ and $\nabla\times\uv = (0,0,2\Omega(\zeta))$, directed along the axis.

\begin{figure}
\includegraphics[scale=0.6]{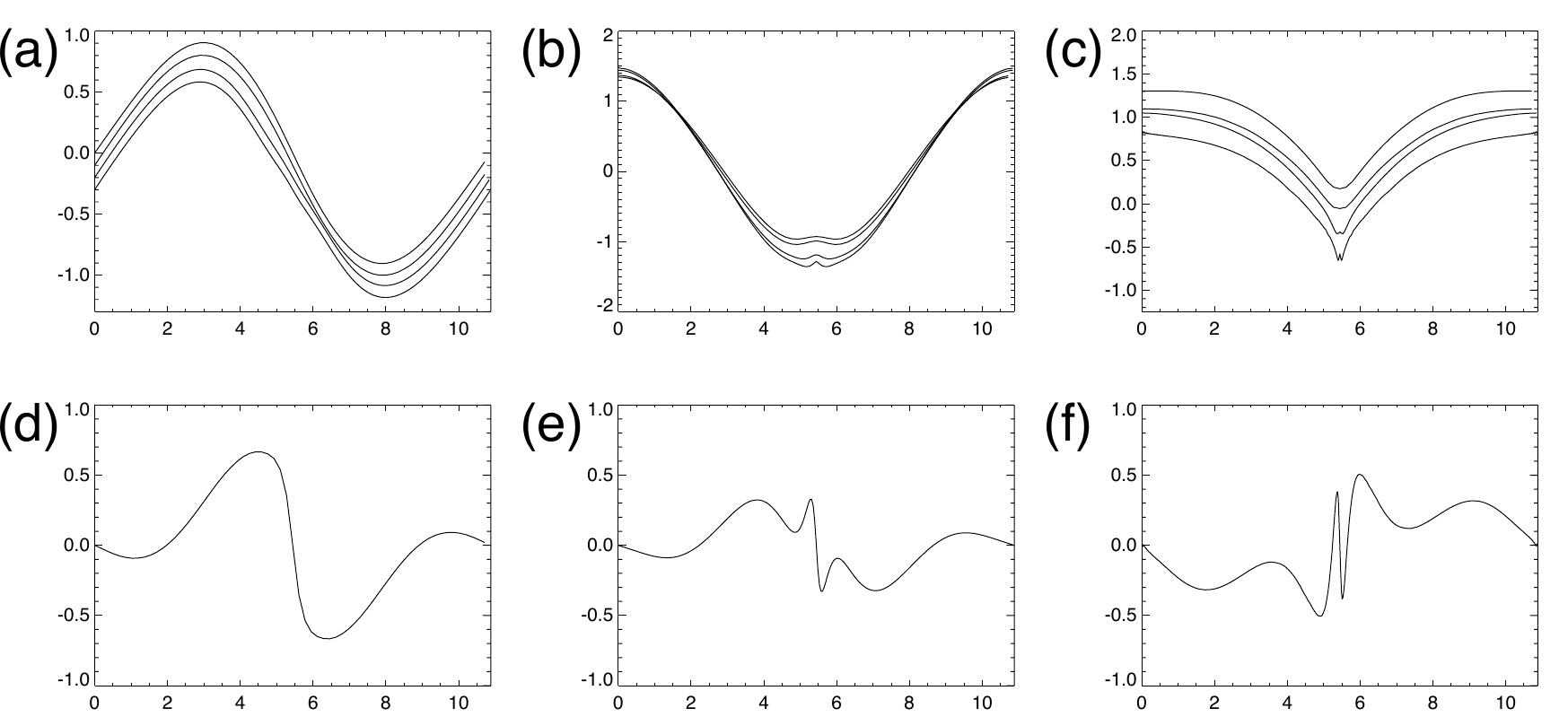}
%\vspace{5cm}
\caption{Plot of components of fields against $\zeta$ for runs with $\eps=0.02$, $0.01$, $10^{-3}$ and $10^{-4}$
for
(a) the field $\Lamv_+$, reading down the curves (separated by adding $0$, $-0.1$, $-0.2$, etc.),
(b) the field $\nabla\times\Lamv_+$, reading down the curves
(c) the field $\eps^{-1/2} \nabla\times \Lamv_-$, reading down the curves (separated by adding $0$, $-0.25$, $-0.5$, etc.)
(d,e,f) the field $\eps^{-1}\Lamv_-$ for $\eps$ equal to (d) $0.01$, (e) $10^{-3}$ and (f) $10^{-4}$.   }
\label{figplotsep}
\end{figure}

Now in the nonlinear, equilibrated regime, the symmetries of the system are observed to be preserved and so the motion near and along the separatrix is given by (\ref{eqsepmotion}) for some functions  $s(\zeta)$ and $\Omega(\zeta)$. These functions characterise aspects of the nonlinear saturation on the separatrices and so of the spiral dissipative structures that form there, visible in figure \ref{figvapor}. We can measure the equivalent functions for any field, and in our runs we find that the traces for $2\uv$, $2\bv$ and $\Lamv_+$ are identical to graphical accuracy. In figure \ref{figplotsep}(a) we show the components of $\Lamv_+$ along the separatrix (separated by constants). This figure in fact depicts two separatrices, the main separatrix from $(0,0,0)$ to $(\pi,\pi,\pi)$ and the next one that continues to $(2\pi,2\pi,2\pi)$, with $0\leq \zeta\leq 2\sqrt{3}\pi$. The components of $\Lamv_+$ show a sinusoidal form in keeping with the property of the equilibrated fields noted by CG, namely%
\footnote{As noted by CG, although this is a good approximation, the error does not go to zero with $\eps$ (e.g., $\lVert | \Lamv_+-\uvKol| \rVert_\infty /   \lVert | \Lamv_+|  \rVert_\infty $ remaining at a level of about 15\% for all runs) and so (\ref{eqallKol}) should not be seen as an asymptotic statement.}
\begin{equation}
\Lamv_+ \simeq \uvKol .
\label{eqallKol}
\end{equation}
There is only slight steepening at $(\pi,\pi,\pi)$ as $\epsilon$ is reduced. Panel \ref{figplotsep}(b) shows traces of the components of $\nabla\times\Lamv_+$ with clear cosine form, in keeping with (\ref{eqallKol}) and (\ref{eqsepmotionfuncs}) but of somewhat enhanced amplitude, and with evidence of some finer scale structure near $(\pi,\pi,\pi)$. These indicate that the approximation (\ref{eqallKol}) is reasonable for the leading  order fields on the separatrices.

The picture is naturally more complicated for the $\Lamv_-$ field, which tends to zero in the limit of small $\eps$. Panels \ref{figplotsep}(d,e,f) plot the components of $\eps^{-1}\Lamv_-$  along the separatrix: there is clear evidence of finer scale oscillations emerging in the limit, but the nature of the limiting distribution is unclear. Panel \ref{figplotsep}(c) shows the fields $\eps^{-1/2}\nabla\times\Lamv_-$ (separated by constants). These show the development of a cusp at $(\pi,\pi,\pi)$, the stagnation point where the two separatrices converge.
In conclusion, the field $\Lamv_-$ on the separatrix scales as $O(\eps)$, but its curl scales as $O(\eps^{1/2})$, giving a natural $O(\eps^{1/2})$ cigar width length scale, confirming results in CG and to be explored further below.

\section{Local behaviour and scaling in the cigars}\label{secPDE}

We now have some knowledge of the local structure of the flow and field on the separatrices, in terms of both the general form it must take, namely (\ref{eqsepmotion}), and the actual behaviour for small values of $\eps$ in figure \ref{figplotsep}. The aim of the present section is to derive the dissipative lengthscale of $\sqrt{\eps}$ noted by CG. Of course we are not able to put together a solution that is complete: the dissipative, cigar-like regions process field that is drawn in, in a spiralling fashion, and then churn it out again. A complete picture would involve matching to the outer region, which is a highly three-dimensional problem, beyond what we can do; nonetheless a local picture gives some information.

\subsection{Uncurling the induction equation}

We start with the formulation in Elsasser variables (\ref{eqElsdef}--\ref{eqElsdiv}) and for brevity set
\begin{equation}
\Lamv \equiv \Lamv_+, \quad
\eps \lamv \equiv \Lamv_-, \quad
p \to \eps p .
\end{equation}
We assume the key scaling of CG that $\lamv=O(1)$, at least in the outer region, which means away from the stagnation points and the separatrices. Without approximation, the steady equations are
\begin{align}
\lamv \cdot\nabla\Lamv & = -\nabla p + \nabla^2 \Lamv +  \fv ,
\label{eqLamv}\\
\Lamv \cdot\nabla\lamv & = -\nabla p +\eps \nabla^2 \lamv +  \fv .
\label{eqlamv}
\end{align}
Note that a straightforward estimate of the width of a diffusive layer based on (\ref{eqlamv}) would suggest an order $\eps$ scaling from balancing $\Lamv\cdot\nabla\lamv\sim\eps\nabla^2\lamv$, but this is too small, as it does not take into account the different scales of variation of $\lamv$ along and across the characteristics of $\Lamv$,  and the following, more delicate argument is needed.

Subtracting (\ref{eqlamv}) from (\ref{eqLamv}) gives an equation equivalent to the induction equation (\ref{eqinduc}),
\begin{equation}
0 = \nabla\times(\lamv\times\Lamv) + \nabla^2 \Lamv - \eps\nabla^2 \lamv,
\end{equation}
which may be uncurled as
\begin{equation}
\nabla a = \lamv \times\Lamv - \nabla\times\Lamv + \eps\nabla\times\lamv,
\label{equncurl}
\end{equation}
where $a(\rv)$ is a scalar field. Taking the divergence gives an elliptic equation for $a$,
\begin{equation}
\nabla^2 a = \nabla \cdot(\lamv\times\Lamv).
\end{equation}

This development can be pursued further, to obtain a general closed but complicated system of scalar PDEs that link the field and flow to the external forcing, as in \cite{Zh09}. However our present aims are more limited: we only need that (\ref{equncurl}) is equivalent to two equations,
\begin{equation}
\Lamv \cdot\nabla a   = -  \Lamv \cdot\nabla\times\Lamv  +  \eps \Lamv\cdot\nabla\times\lamv
\label{eqatransport}
\end{equation}
and
\begin{equation}
\lamv =  c\Lamv + \Lambda^{-2} \Lamv\times( \nabla a  + \nabla\times\Lamv - \eps\nabla\times \lamv) ,
\label{eqlamvac}
\end{equation}
where $c(\rv)$ is another scalar field which obeys
\begin{equation}
\Lamv \cdot\nabla c = -\nabla\cdot[ \Lambda^{-2} \Lamv\times( \nabla a  + \nabla\times\Lamv - \eps\nabla\times \lamv)] ,
\label{eqctransport}
\end{equation}
from requiring that $\nabla\cdot\lamv =0 $. Everything is exact up to this point but we note that this representation will generally break down at isolated points where $\Lamv = 0$.

Now we approximate: first consider an `outer' region, well away from the dissipative, cigar-like structures that lie on the separatrices joining stagnation points.We neglect diffusion in the outer region, the fields having a greater length-scale. The leading order outer problem is obtained by simply setting $\eps=0$ in the above equations (\ref{eqatransport}--\ref{eqctransport}), leaving a pair of quasi-linear equations for $a$ and $c$ giving transport along characteristics of $\Lamv$, namely
\begin{equation}
\Lamv \cdot\nabla a    = -  \Lamv \cdot\nabla\times\Lamv   ,
\label{eqatransport0}
\end{equation}
\begin{equation}
\Lamv \cdot\nabla c  = -\nabla\cdot[ \Lambda^{-2} \Lamv\times( \nabla a  + \nabla\times\Lamv)] ,
\label{eqctransport0}
\end{equation}
together with an equation that then reconstructs $\lamv$, from (\ref{eqlamvac}), which we write as a sum of three terms,
\begin{equation}
\lamv  =  \lamv_c  + \lamv_a + \lamv_{\Lamv},
\end{equation}
with
\begin{equation}
\lamv_c  =  c\Lamv , \quad
\lamv_a  =  \Lambda^{-2} \Lamv\times \nabla a  , \quad
\lamv_{\Lamv} =  \Lambda^{-2} \Lamv\times (\nabla\times\Lamv) .
\label{eqlamvac0}
\end{equation}

Finally for this section, we note that in the outer region, $\lamv_c$ can be
calculated explicitly in terms of $\Lamv$ and $a$. Substitution of \rf{eqlamvac0} into \rf{eqlamv}, where the diffusive term involving $\eps$ is neglected, yields
\begin{equation}
\Rot\left[(\Lamv\cdot\nabla)\left(c\Lamv+\Lambda^{-2}\Lamv\times
(\nabla a+\Rot\Lamv)\right)\right]=\Rot\fv.
\end{equation}
By virtue of \rf{eqctransport0}, this equation takes the form
\begin{equation}
\nabla c\times(\Lamv\cdot\nabla)\Lamv+c\Rot((\Lamv\cdot\nabla)\Lamv)=\Rot\Fv,
\label{form}\end{equation}
where
\begin{equation}
\Fv\equiv\Lamv\Div[\Lambda^{-2}\Lamv\times(\nabla a+\Rot\Lamv)]
-(\Lamv\cdot\nabla)[\Lambda^{-2}\Lamv\times(\nabla a+\Rot\Lamv)]+\fv.
\end{equation}
Scalar multiplication of \rf{form} by $(\Lamv\cdot\nabla)\Lamv$ yields
\begin{equation}
c=\frac{(\Rot\Fv)\cdot(\Lamv\cdot\nabla\Lamv)}{[\Rot(\Lamv\cdot\nabla\Lamv)]\cdot[\Lamv\cdot\nabla\Lamv]}.
\label{cvalue}
\end{equation}
Thus singularities of $c$ can arise, where the helicity type term involving
$(\Lamv\cdot\nabla)\Lamv$ (i.e., the denominator in \rf{cvalue}) vanishes.
%(from this perspective, ordinary null points of $\Lamv$ or
%$(\Lamv\cdot\nabla)\Lamv$ are not problematic); note, however, that
%the product $c\Lamv$, and not the quantity $c$ on its own, is relevant.

\subsection{Field in the outer region, near the main separatrix}

In the outer region, as the main separatrix is approached, it is observed that the field  $\Lamv$ is relatively smooth, as seen from the numerical simulations of CG, and also in view of the leading order Laplacian in (\ref{eqLamv}), whereas $\lamv$ develops fine scales. Using the formulation in (\ref{eqsepmotion}) we from now on define $s(\zeta)$ and $\Omega(\zeta)$ by
\begin{equation}
\Lamv = (-s'(\zeta)\rho, \Omega(\zeta)\rho, 2s(\zeta) ) + O(\rho^2) ,
\label{eqLamvsep}
\end{equation}
in cylindrical polar coordinates $(\rho, \theta,\zeta)$ defined in (\ref{eqcoorddef}) and below. Here the functions $2s(\zeta)/\sqrt{3}$ and $2\Omega(\zeta)\sqrt{3}$ defined for $\Lamv$  are shown in figure \ref{figplotsep}(a,b) and are not known analytically; nonetheless their functional form is similar to that of the Kolmogorov flow (\ref{eqsepmotionfuncs})

To understand the diffusive $O(\eps^{1/2})$ scaling in the cigars and to determine something of the local structure of the fine-scaled $\lamv$ field the following strategy is adopted: solve the equations (\ref{eqatransport0}) and (\ref{eqctransport0}) by integrating along characteristics of $\Lamv$ given locally by (\ref{eqLamvsep}) and reconstruct $\lamv$ via (\ref{eqlamvac0}). As the characteristics of $\Lamv$ approach the origin and are squeezed along the outgoing separatrix, given by $\zeta=O(1)$, $\rho=0$, high gradients build up and the terms in $\eps$ that were earlier neglected increase: when these come into balance with the terms we have retained, we reach the scale at which diffusive effects become important, fixing the width of the dissipative regions.

\begin{figure}
\includegraphics[scale=0.5]{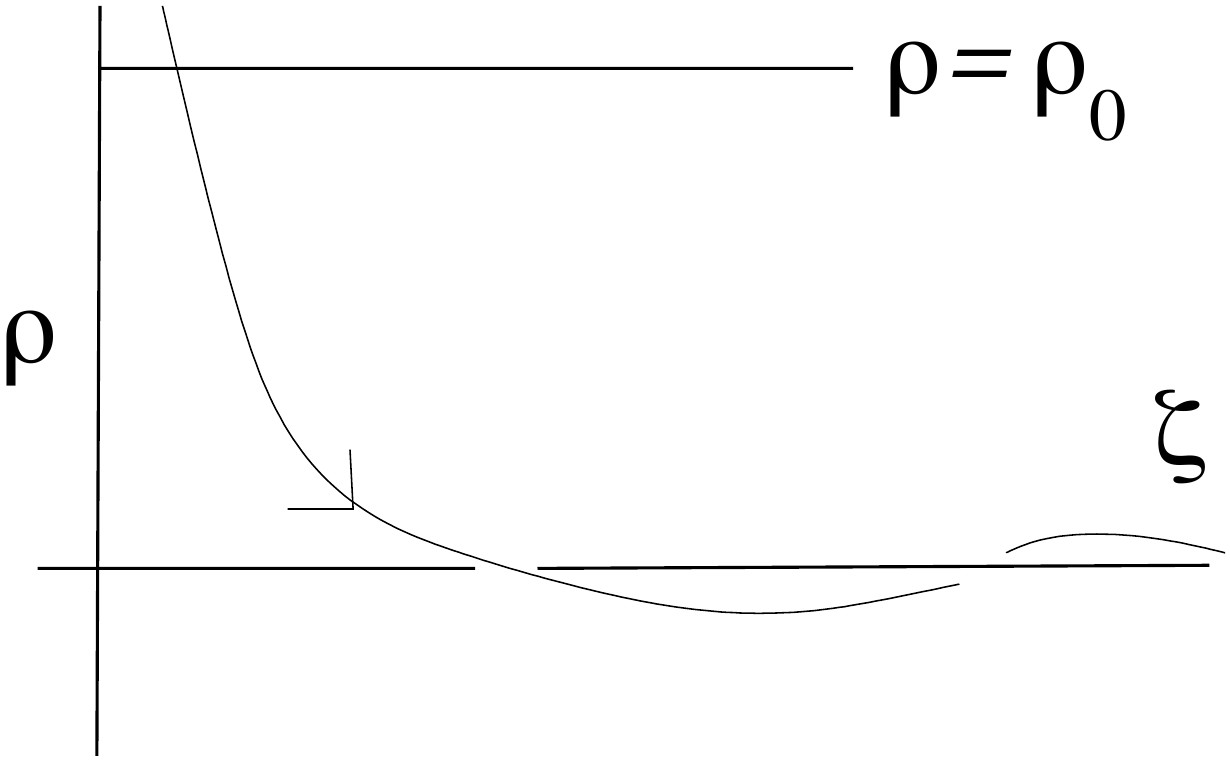}
%\vspace{5cm}
\caption{Schematic figure showing the flow in the $(\rho, \theta, \zeta)$ coordinates.}
% and the region $\zeta=O(1)$ and $\rho\ll1$.
\label{figschem}
\end{figure}
There are two problems with this approach: first that the incoming values of $a$ and $c$ are determined by the outer solution and links to other cigars: as this is beyond what can be addressed analytically, unknown functions have to be introduced. Secondly, even with the simplified, general local form (\ref{eqLamvsep}), analytical calculations rapidly become unwieldy. The first problem will remain with us, but to ameliorate the second problem we simplify further and consider only the motion near to the origin, in which will simply take the field $\Lamv$ to be, exactly,
\begin{equation}
\Lamv = (- \sigma \rho, \omega\rho, 2\sigma \zeta) ,
\label{eqLamvstag}
\end{equation}
in the local cylindrical polar coordinate system $(\rho,\phi,\zeta)$. Here $\sigma$ and $\omega$ are taken as constants, which we may identify as
\begin{equation}
\sigma = s'(0),
\quad
\omega = \Omega(0) .
\end{equation}
We also note from (\ref{eqLamvstag}) that
\begin{equation}
\nabla\times\Lamv = (0,0,2\omega), \quad
\Lambda^{2}  = ( \omega^2+ \sigma^2) \rho^2 + 4\sigma^2\zeta^2 = 4\sigma^2\zeta^2 + O(\rho^2).
\end{equation}

Our strategy now is to solve the outer, diffusionless equations (\ref{eqatransport0}--\ref{eqctransport0}) for transport of $a$ and $c$ for the simplified form (\ref{eqLamvstag}) of $\Lamv$. This is done exactly, but then to see how large the neglected, diffusion terms are, we approximate by taking $\zeta=O(1)$ but $\rho\ll1$, so our results are valid in the outer region, near to the origin, on the outward-going separatrix, as depicted schematically in figure \ref{figschem}. Of course, by the time $\zeta=O(1)$ we are, strictly speaking, away from the stagnation point at the origin and the form (\ref{eqLamvstag}) that we are using no longer applies. However the above form is sufficient to obtain the overall structure of the outer solution as the separatrix is approached, together with the scaling of the diffusive layer width.

Equation (\ref{eqatransport0}) becomes
\begin{equation}
\Lamv \cdot\nabla a = -4\omega\sigma\zeta,
\label{eqatravel}
\end{equation}
and letting $t$ be a time parameter along characteristics, we integrate this in the standard way, with
\begin{equation}
\dot{\rho} = -\sigma\rho, \quad
\dot{\theta} = \omega, \quad
\dot{\zeta} = 2\sigma\zeta, \quad
\dot{a}= - 4\omega\sigma\zeta,
\end{equation}
and the solution in terms of initial conditions on a characteristic,
\begin{equation}
\rho  = \rho_0 e^{-\sigma t} , \quad
\theta  = \theta_0 + \omega t, \quad
\zeta  = \zeta_0 e^{2\sigma t} , \quad
a = a_0 + 2\omega\zeta_0 (1-e^{2\sigma t}) .
\end{equation}
If we suppose that we specify the incoming values of $a$ on a surface $\rho=\rho_0>0$ (see figure \ref{figschem}) with
\begin{equation}
a_0 = a(\rho_0, \theta_0, \zeta_0) = A(\theta_0, \zeta_0)
\end{equation}
at $t=0$, then we have the solution:
\begin{equation}
a(\rho, \theta, \zeta) = A[\theta + \sigma^{-1} \omega\log(\rho/\rho_0), \zeta\rho^2/\rho_0^2] + 2\omega\zeta( \rho^2/\rho_0^2 -1 ).
\label{eqasolve}
\end{equation}
Here $A$ gives the form of the field being carried in from the outer region, and we do not know much about it, except that it has 3-fold rotational symmetry (see figure \ref{figvapor} and figure 13 of CG). It is perhaps helpful to think of $A$ as being some function of order unity with the appropriate symmetry, for example $A = A_0 + A_3 \cos3\theta$.

Given $a$ we can now reconstruct the appropriate part of $\lamv$ in (\ref{eqlamvac0}). We have
\begin{align}
\nabla a &= ( \omega\sigma^{-1} \rho^{-1} A_\theta
+ 2\zeta\rho \rho_0^{-2} (A_\zeta + 2\omega) ,
\rho^{-1}A_\theta, \rho^2\rho_0^{-2} A_\zeta + 2\omega(\rho^2\rho_0^{-2} -1 ) ) \notag\\
& = \sigma^{-1} \rho^{-1}  (\omega A_\theta , \sigma A_\theta, -2\omega\sigma\rho)+O(\rho) ,
\end{align}
and so
\begin{equation}
\lamv_a \equiv \Lambda^{-2} \Lamv\times\nabla a = 2 \Lambda^{-2}(-\sigma,\omega,0)  \zeta\rho^{-1} A_{\theta} + O(\rho^0),
\end{equation}
as $\rho\to0$, where $A_\theta$ denotes the derivative of $A$ with respect to its first argument. Here we have obtained a component growing as $\rho^{-1}$ which arises because of the incoming values of $A$ on different characteristics being squeezed together.

\subsection{The effect of diffusive terms}
\label{ssecdiff}

With this component of $\lamv$ in hand, we can now revisit the diffusive equation (\ref{eqatransport}). We calculate
\begin{equation}
\nabla\times\lamv_a = 2\Lambda^{-2}(\omega^2+\sigma^2)(0,0,\sigma^{-1}) \zeta \rho^{-2} A_{\theta\theta} + O(\rho^{-1}) ,
\end{equation}
and
\begin{equation}
\Lamv \cdot \nabla\times\lamv_a = 4\Lambda^{-2}(\omega^2+\sigma^2) \zeta^2 \rho^{-2} A_{\theta\theta} + O(\rho^{-1}) .
\end{equation}
This now has a $\rho^{-2}$ growth, by virtue of differentiating $A$ again. In equation (\ref{eqatransport}) it is clear that the final $\eps\Lamv \cdot \nabla\times\lamv_a$ term with diffusion will be the same order as the term $\Lamv \cdot \nabla\times\Lamv=4\omega\sigma\zeta$  we originally included, when $\eps\rho^{-2}=O(1)$. This gives the $\rho=O(\sqrt{\eps})$ scaling of the diffusive cigar width. Similarly at these values of $\rho$, in (\ref{eqlamvac}) and (\ref{eqctransport}) the terms  $\eps\nabla\times\lamv_a$ become of similar magnitude to $\nabla\times\Lamv$ (though note here that the $\nabla a$ terms are actually larger in magnitude at this point).

This is the main part of the argument: although we have simplified by focusing solely on $\lamv_a$ in (\ref{eqlamvac0}), consideration of the scalar $c$ and component $\lamv_c$ does not affect the discussion, nor does $\lamv_{\Lamv}$, given straightforwardly by
\begin{equation}
\lamv_{\Lamv} \equiv \Lambda^{-2} \Lamv\times(\nabla\times\Lamv)
= 2 \Lambda^{-2} \omega (\omega, \sigma, 0) \rho ,
\end{equation}
and so is negligible. To check this we now look at the $\eps=0$ equation (\ref{eqctransport0}) for $c$ and the corresponding component $\lamv_c$. After a straightforward calculation (\ref{eqctransport0}) becomes
\begin{equation}
\Lamv\cdot\nabla c = -12 \Lambda^{-4} \sigma (\omega^2+\sigma^2) \zeta A_\theta + O(\rho).
\end{equation}
The key point is that the right-hand side is of order unity as $\rho\to0$, as was the case for $a$ in (\ref{eqatravel}). Thus without solving the equation in detail, it is clear that the solution analogous to (\ref{eqasolve}) for $a$ will take the form
\begin{equation}
c(\rho, \theta, \zeta) = C[\theta + \omega\sigma^{-1} \log(\rho/\rho_0), \zeta\rho^2/\rho_0^2] + C_{\mathrm{PI}}(\rho, \theta,\zeta) ,
\end{equation}
where $C(\theta_0,\zeta_0)$ gives the incoming values of $c$ on the surface $\rho=\rho_0$, as before and the particular integral $C_{\mathrm{PI}}$ involves $A$ but is of order unity as $\rho\to 0 $.

Now when we reconstruct $\lamv$ via (\ref{eqlamvac0}), the component $c\Lamv = O(1)$ along streamlines will be subdominant to the component $\Lamv \times\nabla a = O(\rho^{-1})$, the inverse power of $\rho$ arising from taking the gradient. Thus our focus on $\lamv_a$ in the above discussion of the diffusive breakdown of the outer solution is justified and we have
\begin{equation}
\lamv
 = \lamv_a + O(1)
 = 2\Lambda^{-2} (-\sigma,\omega,0) \zeta\rho^{-1} A_\theta + O(1) ,
\end{equation}
as $\rho\to0$ on the outgoing separatrix.
As a by-product of our calculations we observe that the small-scale field $\lamv$ will show components $\lamv_a$ perpendicular to streamlines that diverge as $\rho^{-1}$ as the separatrix is approached from the outer solution. These will peak at levels $\lamv = O(\eps^{-1/2})$ when diffusive suppression begins to occur at scales $\rho=\sqrt{\eps}$. This is in keeping with the scalings seen by CG, who note that  $\Lamv_- =\eps\lamv= O(\sqrt{\eps})$ near the separatrices (their section 3.2.1, figures 13 and 16). In view of the $\cos3\theta$ dependence of the leading field identified here, this component must go to zero on the axis itself and is presumably strongly suppressed by diffusion. Thus we cannot make a detailed link with figure \ref{figplotsep}: the field here originates with the mean component of $A$, independent of $\theta$, for which the onset of diffusion will be delayed until smaller values of $\rho$. This also presumably explains the structure seen in figure \ref{figvapor} (most clearly in (b)) or figure 13 of CG, with three incoming sheets of field merging in an axisymmetric `collar' at smaller values of $\rho$. In this way, there could be several nested boundary layers along the separatrices in the limit $\eps\to0$.

\section{Existence of weak steady-state solutions}
\label{maths}

We consider now the system of equations \rf{eqLamv} and \rf{eqlamv}
in Elsasser variables, together with the solenoidality conditions
\begin{equation}
\Div\lamv=\Div\Lamv=0.
\label{lamsol}\end{equation}
In this section we define weak solutions to these equations and formally prove
their existence, adapting the approach of \cite{Lad69}. In the next two sections
we will show that the weak solutions are classical
smooth functions, satisfying the equations at any point in space.

We start by recalling some definitions. Consider the class of functions
whose domain is the periodicity cell $\Tbb^3 \equiv[0,2\pi]^3$. The norm
in the Lebesgue space $L_p(\Tbb^3)$ is defined, for $p\ge 1$, as
\begin{equation}
\|\Phiv\|_p\equiv\left(\int_{\Tbb^3}|\Phiv|^p\dx\right)^{1/p}.\end{equation}
Since in the above-mentioned class $\Ic-\nabla^2$ is a positively defined
self-adjoint operator
(where $\Ic$ is the identity), whose eigenfunctions are Fourier harmonics, we
can define in the usual way the powers $(\Ic-\nabla^2)^\alpha$ for an arbitrary
real $\alpha$, by considering Fourier series. For $\rv\in R^3$ and any
\begin{equation}
\Phiv=\sum_{\nv\ne 0}\Phiv_{\nv}\,e^{i\nv\cdot\rv},\end{equation}
\begin{equation}
(\Ic-\nabla^2)^\alpha\Phiv\equiv
\sum_{\nv\ne 0}(1+|\nv|^2)^\alpha\Phiv_{\nv}\,e^{i\nv\cdot\rv}.\end{equation}
The Sobolev space $W^s_p(\Tbb^3)$ is defined for $p\ge 1$, as the closure
in the norm
\begin{equation}
\|\Phiv\|_{s,p}\equiv\|(\Ic-\nabla^2)^{s/2}\Phiv\|_p\label{nWsp}\end{equation}
of the set of infinitely smooth periodic functions, whose domain is $\Tbb^3$.
(Evidently, $L_p(\Tbb^3)=W^0_p(\Tbb^3)$.) We will work in the subspace of
zero-mean vector fields, in which the operator $-\nabla^2$ can be used instead
of $\Ic-\nabla^2$ in these definitions. In particular, we define (without
introducing a new notation) a norm, equivalent to \rf{nWsp}, in the subspace of
zero-mean fields in $W^s_p(\Tbb^3)$ as
\begin{equation}
\|\Phiv\|_{s,p}\equiv\|(-\nabla^2)^{s/2}\Phiv\|_p.\end{equation}
Since the Laplacian is a self-adjoint operator, in the important particular
case $p=2$ this implies
\begin{equation}
\|\Phiv\|_{s,2}^2=\int_{\Tbb^3}\Phiv\cdot(-\nabla^2)^s\Phiv\dx.\label{goodno}\end{equation}
We will employ the following:

{\bf Embedding theorem} (see \cite{BL76}, \cite{Tay81} and references therein).

(i) For $s>N/p$, $W^s_p(\Tbb^3)\subset C(\Tbb^3)$.

(ii) For $0<s<N/p$ and $q=Np/(N-ps)$, $W^s_p(\Tbb^3)\subset L_q(\Tbb^3)$
(in particular, $\|\Phiv\|_q\le C_{s,p}\|\Phiv\|_{s,p}$).

We will show in the remainder of this section that for any space-periodic
forcing $\fv$ from the Lebesgue space $L_2(\Tbb^3)$, the system of equations
\rf{eqLamv}, \rf{eqlamv} and \rf{lamsol} has at least one weak space-periodic
solution from the Sobolev space $W^1_2(\Tbb^3)$. The assumption that the box
of periodicity is the cube $\Tbb^3\equiv[0,2\pi]^3$ is technical: our arguments
can be repeated almost literally for the case of an arbitrary parallelepiped of
periodicity. Note that in this and the following sections we do not restrict
ourselves to the Kolmogorov forcing \rf{eqSvdef}; higher regularity of $\fv$
will be required in \S\ref{smooth}.

Consider then, the set of infinitely smooth solenoidal zero-mean periodic functions,
whose domain is the periodicity cell $\Tbb^3\equiv[0,2\pi]^3$, and denote by
$\Hc$ its closure in the Sobolev space $W_2^1(\Tbb^3)$. A pair of vector fields
$\Lamv\in\Hc,\ \lamv\in\Hc$ is {\it a weak solution} to the system \rf{eqLamv},
\rf{eqlamv} and \rf{lamsol}, if the integral identities
\begin{equation}
\int_{\Tbb^3}\left(\,\sum_{k=1}^3 \frac{\partial\Lamv}{\partial x_k}\cdot
\frac{\partial\Phiv}{\partial x_k}+((\lamv\cdot\nabla)\Lamv-\fv)\cdot\Phiv\right)\dx=0
\label{Clamint}\end{equation}
and
\begin{equation}
\int_{\Tbb^3}\left(\eps\sum_{k=1}^3 \frac{\partial\lamv}{\partial x_k}\cdot
\frac{\partial\Phiv}{\partial x_k}+((\Lamv\cdot\nabla)\lamv-\fv)\cdot\Phiv\right)\dx=0
\label{Slamint}\end{equation}
hold true for any vector field $\Phiv\in\Hc$. (If $\Lamv$ and $\lamv$ are
smooth, these identities immediately follow from \rf{eqLamv} and \rf{eqlamv}.)
By H\"older's inequality and the embedding theorem, for any
function $f\in W^1_2(\Tbb^3)$,
\begin{equation}
\|f\|_4\le\|f\|_2^{1/4}\|f\|_6^{3/4}\le C_1\|f\|_2^{1/4}\|f\|_{1,2}^{3/4}\le C_1\|f\|_{1,2},
\label{fbound}\end{equation}
where $C_1$ is a constant independent of $f$. Consequently, the
Cauchy--Bunya\-kowsky--Schwarz inequality implies that
the integrals involving nonlinear terms admit the bounds
\begin{align}
\left|\int_{\Tbb^3}((\lamv\cdot\nabla)\Lamv)\cdot\Phiv\dx\right| & =
\left|\int_{\Tbb^3}\sum_{k=1}^3\lambda_k\Lamv\cdot \frac{\partial\Phiv}{\partial x_k}\dx\right|
\label{Llbound}
\\
& \le\sum_{j=1}^3\sum_{k=1}^3\|\lambda_k\|_4\|\Lambda_j\|_4
\left\|\frac{\partial\Phi_j}{\partial x_k}\right\|_2\le
C_2\|\lamv\|_{1,2}\|\Lamv\|_{1,2}\|\Phiv\|_{1,2},
\notag
\end{align}
$C_2$ being a constant independent of $\lamv,\Lamv$ and $\Phiv$, and similarly
\begin{equation}
\left|\int_{\Tbb^3}((\Lamv\cdot\nabla)\lamv)\cdot\Phiv\dx\right|\le
C_2\|\lamv\|_{1,2}\|\Lamv\|_{1,2}\|\Phiv\|_{1,2}.\end{equation}
Thus the integrals are well-defined.

Consider the scalar product in $\Hc$
\begin{equation}
[\Phiv_1,\Phiv_2]\equiv\int_{\Tbb^3}\sum_{k=1}^3
\frac{\partial\Phiv_1}{\partial x_k}\cdot\frac{\partial\Phiv_2}{\partial x_k}\dx.%
\label{dotp}\end{equation}
Integrating by parts we recast the identities \rf{Clamint} and \rf{Slamint}
in an alternative form involving the scalar product \rf{dotp}:
\begin{equation}
[\Lamv-\Ac(\lamv,\Lamv)-\fvt,\Phiv]=0,
\label{Ldot}\end{equation}
and
\begin{equation}
[\eps\lamv-\Ac(\Lamv,\lamv)-\fvt,\Phiv]=0.
\label{ldot}\end{equation}
Here
\begin{equation}
\fvt=-(\nabla^2)^{-1}\fv,\end{equation}
$(\nabla^2)^{-1}$ denoting, as usual, the inverse Laplacian, and
\begin{equation}
\Ac(\lamv,\Lamv)\equiv(\nabla^2)^{-1}\Pc((\lamv\cdot\nabla)\Lamv)%
\end{equation}
is a bilinear operator, where $\Pc$ is the projection onto the subspace of
solenoidal vector fields. (In fact, for the Kolmogorov forcing $\fvt=\fv$, but
in what follows we do not employ this equality.)

Using \rf{goodno} for $s=1$, we find
\begin{align}
\|\Ac(\lamv,\Lamv)\|_{1,2}^2 &=-\int_{\Tbb^3}
\Pc((\lamv\cdot\nabla)\Lamv)\cdot(\nabla^2)^{-1}\Pc((\lamv\cdot\nabla)\Lamv)\dx%
\notag
\\
& =\sum_{j=1}^3\sum_{k=1}^3\int_{\Tbb^3}\lambda_k\Lamv\cdot(\nabla^2)^{-1}
\Pc\left[\frac{\partial^2}{\partial x_j\partial x_k}(\lambda_j\Lamv)\right]\dx.
\label{ocenka}\end{align}
For any
\begin{equation}
\Phiv=\sum_{\nv \ne 0}\Phiv_{\nv} \,e^{i\nv\cdot\rv},\end{equation}
\begin{equation}
(\nabla^2)^{-1}\Pc \left[ \frac{\partial^2}{\partial x_j\partial x_k}\Phiv \right ]
=\sum_{\nv\ne 0}\left(\Phiv_{\nv}- \frac{\Phiv_{\nv} \cdot\nv}{|\nv|^2}\right)
\frac{n_jn_k}{|\nv|^2}\,e^{i\nv\cdot\rv},%
\end{equation}

and therefore
\begin{equation}
\left\|(\nabla^2)^{-1}\Pc\left[\frac{\partial^2}{\partial x_j\partial x_k}\Phiv\right]\right\|_2
\le\|\Phiv\|_2.\end{equation}
Now we develop \rf{ocenka}, using H\"older's inequality and the embedding theorem,
\begin{equation}
\|\Ac(\lamv,\Lamv)\|_{1,2}^2\le\sum_{j=1}^3\sum_{k=1}^3\sum_{l=1}^3
\|\lambda_k\|_4\|\Lambda_l\|_4\|\lambda_j\Lamv\|_2\le C_3\|\lamv\|_{1,2}^2\|\Lamv\|_{1,2}^2,
\end{equation}
which shows that $\Ac:\Hc\otimes\Hc\to\Hc$.

Thus, we have shown that for $\Lamv\in\Hc$ and $\lamv\in\Hc$
the first factors in the scalar products in the right-hand sides of \rf{Ldot}
and \rf{ldot} belong to $\Hc$. Since smooth vector fields are dense
in $\Hc$ in the norm induced by the scalar product $[\cdot,\cdot]$,
\rf{Ldot} and \rf{ldot} are equivalent to equations
\begin{equation}
\Lamv-\Ac(\lamv,\Lamv)-\fvt={\bf 0}
\label{Leqn}\end{equation}
and
\begin{equation}
\lamv- {\eps^{-1}}(\Ac(\Lamv,\lamv)+\fvt)={\bf 0},
\label{leqn}\end{equation}
respectively, understood as equalities in $\Hc$.

The existence of solutions to the system \rf{Leqn}, \rf{leqn} is guaranteed by the
Leray--Schauder principle (see \cite{LeSch} and \cite{Lad69}) under two conditions:

(i) The operator $\Bc:\Hc\otimes\Hc\to\Hc\otimes\Hc$ defined as
\begin{equation}
\Bc(\Lamv,\lamv)=(\Ac(\lamv,\Lamv),\Ac(\Lamv,\lamv)/\eps)\end{equation}
is compact, i.e. $\Bc(\Lamv_n,\lamv_n)$ is a strongly converging
sequence in $\Hc\otimes\Hc$ for any sequence $(\Lamv_n,\lamv_n)$
weakly converging in $\Hc\otimes\Hc$.

(ii) Any solution to the set of equations
\begin{equation}
\Lamv-\mu(\Ac(\lamv,\Lamv)+\fvt)= 0,\qquad
\lamv-{\mu}{\eps^{-1}}(\Ac(\Lamv,\lamv)+\fvt)= 0
\label{leray}\end{equation}
belongs to a ball in $\Hc\otimes\Hc$ of a radius independent of $\mu$
for $0\le\mu\le1$.

The proof of (i) relies on the embedding theorem for Sobolev spaces, whereby
the embedding $W^1_2(\Tbb^3)\to L_q(\Tbb^3)$ is compact for $q<6$, i.e.,
$\|\lamv^n-\lamv^m\|_q+\|\Lamv^n-\Lamv^m\|_q\to0$ for $m,n\to\infty$,
for any sequence $(\lamv^n,\Lamv^n)$ weakly converging in
$W^1_2(\Tbb^3)\otimes W^1_2(\Tbb^3)$. It is enough to prove that
$\Ac(\lamv^n,\Lamv^n)$ converges strongly in $\Hc$. For any $\Phiv\in\Hc$,
\begin{equation}
[\Ac(\lamv^n,\Lamv^n)-\Ac(\lamv^m,\Lamv^m),\Phiv]
=\int_{\Tbb^3}\sum_{k=1}^3(\lambda_k^n\Lamv^n-\lambda_k^m\Lamv^m)
\cdot\frac{\partial\Phiv}{\partial x_k}\dx\end{equation}
\begin{equation}
=\int_{\Tbb^3}\sum_{k=1}^3\lambda_k^n(\Lamv^n-\Lamv^m)
\cdot\frac{\partial\Phiv}{\partial x_k}\dx+
\int_{\Tbb^3}\sum_{k=1}^3(\lambda_k^n-\lambda_k^m)\Lamv^m
\cdot\frac{\partial\Phiv}{\partial x_k}\dx.\end{equation}
Hence, by the same arguments as were used to derive \rf{Llbound}, we obtain
\begin{equation}
\!\!\!\!\!\!\!\!\!    \|[\Ac(\lamv^n,\Lamv^n)-\Ac(\lamv^m,\Lamv^m),\Phiv]\|_{1,2}\le
C_4(\|\lamv^n\|_{1,2}\|\Lamv^n-\Lamv^m\|_4
+\|\lamv^n-\lamv^m\|_4\|\Lamv^m\|_{1,2})\|\Phiv\|_{1,2}.
\end{equation}
Here $\Phiv\in\Hc$ is arbitrary; letting
$\Phiv=\Ac(\lamv^n,\Lamv^n)-\Ac(\lamv^m,\Lamv^m)$,
from this inequality we deduce
\begin{equation}
\|\Ac(\lamv^n,\Lamv^n)-\Ac(\lamv^m,\Lamv^m)\|_{1,2}\le
C_5(\|\Lamv^n-\Lamv^m\|_4+\|\lamv^n-\lamv^m\|_4),%
\end{equation}
where the constant $C_5$ is independent of $m$ and $n$, since weak convergence
of $(\lamv^n,\Lamv^n)$ in $W^1_2(\Tbb^3)\otimes W^1_2(\Tbb^3)$ implies the uniform
boundedness of $\|\lamv^n\|_{1,2}$ and $\|\Lamv^n\|_{1,2}$. Thus we have established
that $\|\Ac(\lamv^n,\Lamv^n)-\Ac(\lamv^m,\Lamv^m)\|_{1,2}\to0$ for
$m,n\to\infty$, as desired.

To prove (ii), we consider the problem \rf{leray} in the form
of integral equations, analogous to \rf{Clamint} and \rf{Slamint},
\begin{equation}
\int_{\Tbb^3}\left(\,\sum_{k=1}^3 \frac{\partial\Lamv}{\partial x_k}\cdot
\frac{\partial\Phiv}{\partial x_k}+\mu((\lamv\cdot\nabla)\Lamv-\fv)\cdot\Phiv\right)\dx=0,
\label{Cleray}\end{equation}
\begin{equation}
\int_{\Tbb^3}\left(\,\sum_{k=1}^3\frac{\partial\lamv}{\partial x_k}\cdot
\frac{\partial\Phiv}{\partial x_k}+\frac{\mu}{\eps}
((\Lamv\cdot\nabla)\lamv-\fv)\cdot\Phiv\right)\dx=0,
\label{Sleray}\end{equation}
which are satisfied for any $\Phiv\in\Hc$.
Let $\Phiv=\Lamv$ in \rf{Cleray} and $\Phiv=\lamv$ in \rf{Sleray}.
Due to solenoidality of $\lamv$ and $\Lamv$ the nonlinear terms vanish,
and we find from the identities \rf{Cleray} and \rf{Sleray}
\begin{equation}
[\Lamv,\Lamv]=\mu\int_{\Tbb^3}\fv\cdot\Lamv\dx\le\mu C_6\|\Lamv\|_2\|\fv\|_2
\quad\Rightarrow\quad\|\Lamv\|_{1,2}\le\mu C_7\|\fv\|_2,%
\end{equation}
\begin{equation}
[\lamv,\lamv]={\mu}{\eps^{-1}}\int_{\Tbb^3}\fv\cdot\lamv\dx\le {\mu}{\eps^{-1}}
C_6\|\lamv\|_2\|\fv\|_2 \quad\Rightarrow\quad\|\lamv\|_{1,2}\le {\mu}{\eps^{-1}}C_7\|\fv\|_2,%
\end{equation}
since the norm, induced by the scalar product $[\cdot,\cdot]$ in $\Hc$,
is equivalent to the norm \rf{goodno} in $W^1_2(\Tbb^3)$.
These inequalities establish the existence of a weak solution $\Lamv\in\Hc$,
$\lamv\in\Hc$ to the problem \rf{eqLamv}, \rf{eqlamv} and \rf{lamsol},
admitting the bounds
\begin{equation}
\|\Lamv\|_{1,2}\le C_7\|\fv\|_2,\qquad\|\lamv\|_{1,2}\le C_7{\eps^{-1}}\|\fv\|_2.
\label{L2bounds}\end{equation}
Here the constant $C_7$ is absolute: it is independent of the solution
$\Lamv$ and $\lamv$, the forcing $\fv$, and the parameter $\eps$.

\section{Bounds for weak solutions in $W^2_2(\Tbb^3)$}\label{bounds}

In this section we obtain bounds for the norms of the weak solution, whose
existence we have established in the previous section, in the Sobolev spaces
$W^{5/4}_2(\Tbb^3)$ and $W^2_2(\Tbb^3)$. The forcing $\fv$ is assumed here
to belong to the Lebesgue space $L_2(\Tbb^3)$, as in the previous section.

Consider the Fourier series
\begin{equation}
\Lamv=\sum_{\nv\ne 0} \Lamv_{\nv}\,e^{i\nv\cdot\rv},\end{equation}
and smooth vector fields
\begin{equation}
\Lamv^M=\sum_{{\nv\ne 0,\ |\nv|\le}M}\Lamv_{\nv}\,e^{i\nv\cdot\rv}\in\Hc.\end{equation}
Scalar multiplying in $L_2(\Tbb^3)$ \rf{Leqn} by $(-\nabla^2)^{5/4}\Lamv^M\in\Hc$,
using self-adjointness of the Laplacian, solenoidality of $\Lamv$ and hence of
$\Lamv^M$, and orthogonality of potential and solenoidal fields in $L_2(\Tbb^3)$,
we obtain
\begin{align}
\int_{\Tbb^3}\Lamv\cdot(-\nabla^2)^{5/4}\Lamv^M\dx  & +
\int_{\Tbb^3}(\lamv\cdot\nabla)\Lamv\cdot(-\nabla^2)^{1/4}\Lamv^M\dx
\notag
\\
& =\int_{\Tbb^3}\fvt\cdot(-\nabla^2)^{5/4}\Lamv^M\dx.
\label{label}\end{align}
Note that
\begin{equation}
\|(-\nabla^2)^{1/4}((-\nabla^2)^{1/4}\Lamv^M)\|_2=\|\Lamv^M\|_{1,2}\le\|\Lamv\|_{1,2},\end{equation}
and hence $(-\nabla^2)^{1/4}\Lamv^M\in W^{1/2}_2(\Tbb^3)$, and by part (ii) of the Theorem
\begin{equation}
\|(-\nabla^2)^{1/4}\Lamv^M\|_3\le C_{1/2,2}\|\Lamv\|_{1,2}.\end{equation}
This, together with \rf{L2bounds}, implies a bound for the first integral
\begin{align}
\left|\int_{\Tbb^3}(\lamv\cdot\nabla)\Lamv\cdot(-\nabla^2)^{1/4}\Lamv^M\dx\right|
 & \le\|\lamv\|_6\|\nabla\Lamv\|_2\|(-\nabla^2)^{1/4}\Lamv^M\|_3
\notag
\\
 & \le C_{1/2,2}\|\Lamv\|_{1,2}^2\|\lamv\|_{1,2}\le C_8/\eps.
 \end{align}
Also,
\begin{equation}
\left|\int_{\Tbb^3}\fvt\cdot(-\nabla^2)^{5/4}\Lamv^M\dx\right|
\le\|\Lamv\|_{1,2}\|(-\nabla^2)^{3/4}\fvt\|_2\le C_9,\end{equation}
and hence we find from the identity \rf{label}
\begin{equation}
\|(-\nabla^2)^{5/8}\Lamv\|_2^2=\sup_M
\int_{\Tbb^3}\Lamv\cdot(-\nabla^2)^{5/4}\Lamv^M\dx\le C_{10}\eps^{-1}.
\label{norm1}\end{equation}
Similarly, \rf{leqn} yields
\begin{equation}
\|(-\nabla^2)^{5/8}\lamv\|_2^2\le C_{10}\eps^{-3}.
\label{norm2}\end{equation}
(The constant $C_{10}$ in \rf{norm1} and \rf{norm2} is independent of $\eps\le1$,
but depends on the norm $\|\fv\|_2$ of the forcing $\fv\in L_2(\Tbb^3)$.)
We have therefore demonstrated that
$\lamv\in W^{5/4}_2(\Tbb^3)$ and $\Lamv\in W^{5/4}_2(\Tbb^3)$.

Consequently, $(-\nabla^2)^{1/2}\Lamv^M\in W^{1/4}_2(\Tbb^3)$ and
$(-\nabla^2)^{1/2}\Lamv^M\in W^{1/4}_2(\Tbb^3)$. Using part (ii) of the theorem,
we find
\begin{equation}
\|\partial\Lamv/\partial x_k\|_{12/5}\le
C_{1/4,2}\|(-\nabla^2)^{1/8}(\partial\Lamv/\partial x_k)\|_2
\le C_{1/4,2}\|(-\nabla^2)^{5/8}\Lamv\|_2
\label{L3norm}
\end{equation}
and
\begin{equation}
\|\Lamv\|_{12}\le C_{5/4,2}\|(-\nabla^2)^{5/8}\Lamv\|_2.
\label{L12norm}
\end{equation}
Similarly,
\begin{equation}
\|\partial\lamv/\partial x_k\|_{12/5}\le C_{1/4,2}\|(-\nabla^2)^{5/8}\lamv\|_2,\quad
\|\lamv\|_{12}\le C_{5/4,2}\|(-\nabla^2)^{5/8}\lamv\|_2.
\label{l3norm}
\end{equation}
Scalar multiplying in $L_2(\Tbb^3)$ \rf{Leqn} by $(-\nabla^2)^2\Lamv^M$, we obtain
\begin{equation}
\|(-\nabla^2)\Lamv^M\|_2^2+\int_{\Tbb^3}(\lamv\cdot\nabla)\Lamv\cdot(-\nabla^2)\Lamv^M\dx
=\int_{\Tbb^3}\fvt\cdot(-\nabla^2)^2\Lamv^M\dx.
\label{integ}
\end{equation}
By H\"older's inequality,
\begin{equation}
\left|\int_{\Tbb^3}(\lamv\cdot\nabla)\Lamv\cdot(-\nabla^2)\Lamv^M\dx\right|
\le\|\lamv\|_{12}\|\nabla\Lamv\|_{12/5}\|(-\nabla^2)\Lamv^M\|_2\end{equation}
and hence from \rf{integ} and (\ref{L3norm}--\ref{l3norm}),
\begin{align}
\|(-\nabla^2)\Lamv^M\|_2^2
\le C_{11} & \|(-\nabla^2)^{5/8}\lamv\|_2^2\|(-\nabla^2)^{5/8}\Lamv\|_2^2
\notag
\\
  + \tfrac{1}{2} & \|(-\nabla^2)\Lamv^M\|_2^2+\|\fv\|_2\|(-\nabla^2)\Lamv\|_2,\end{align}
whereby
\begin{equation}
\|(-\nabla^2)\Lamv\|_2^2=\sup_M\|(-\nabla^2)\Lamv^M\|_2^2\le C_{12}\eps^{-4}.
\label{gapo}\end{equation}
The same operations applied to \rf{leqn} yield
\begin{equation}
\|(-\nabla^2)\lamv\|_2\le C_{12}\eps^{-3}.\label{gap}\end{equation}
Thus we have demonstrated that $\lamv$ and $\Lamv$ belong to $W^2_2(\Tbb^3)$.
The constant $C_{12}$ in \rf{gapo} and \rf{gap} is independent of the solution
$\Lamv$, $\lamv$ and the small parameter $\eps$, but depends on the
norm $\|\fv\|_2$ of the forcing $\fv\in L_2(\Tbb^3)$.

\section{Smoothness of weak solutions}\label{smooth}

Steady-state hydrodynamic and MHD problems are drastically different from
the evolutionary ones in that one can incrementally establish the smoothness
of their solutions together with the derivatives of arbitrary order (provided
the forcing $\fv$ is sufficiently smooth). In this section we use \rf{Leqn}
and \rf{leqn} to show by induction that $\lamv$ and $\Lamv$, whose existence
we have ascertained in \S\ref{maths}, are in fact smooth vector fields
and therefore constitute a classical space-periodic solution to equations
\rf{eqLamv}, \rf{eqlamv} and \rf{lamsol}.

We assume now that $\lamv\in W^{2k}_2(\Tbb^3)$ and $\Lamv\in W^{2k}_2(\Tbb^3)$ (which
is equivalent to $\|(-\nabla^2)^k\lamv\|_2+\|(-\nabla^2)^k\Lamv\|_2<\infty$)
and $\fv\in W^{2k}_2(\Tbb^3)$ for $k\ge1$,
and show that $\lamv\in W^{2k+2}_2(\Tbb^3)$ and $\Lamv\in W^{2k+2}_2(\Tbb^3)$.

Scalar multiplying in $L_2(\Tbb^3)$ \rf{Leqn} by $(-\nabla^2)^{2k+1}\Lamv^M$,
using self-adjointness of the Laplacian, solenoidality of $\Lamv^M$
and orthogonality of potential and solenoidal fields in $L_2(\Tbb^3)$, we obtain
\begin{align}
\int_{\Tbb^3}\Lamv\cdot(-\nabla^2)^{2k+1}\Lamv^M\dx
& +\sum_{j=1}^3\int_{\Tbb^3}\left((-\nabla^2)^{k-1} \frac{\partial}{\partial x_j}
(\lamv\cdot\nabla)\Lamv\right)\!\cdot\!(-\nabla^2)^k \, \frac{\partial\Lamv^M}{\partial x_j}\dx
\notag
\\
& =\int_{\Tbb^3}(-\nabla^2)^{k-1/2}\fv\cdot(-\nabla^2)^{k+1/2}\Lamv^M\dx,
\end{align}
and thus
\begin{align}
\|(-\nabla^2)^{k+1/2}\Lamv^M\|_2^2 & \le\left(\,\sum_{j=1}^3\left\|(-\nabla^2)^{k-1}
\frac{\partial}{\partial x_j}(\lamv\cdot\nabla)\Lamv\right\|_2\right.
\\
& \qquad\qquad\qquad\qquad +\|(-\nabla^2)^{k-1/2}\fv\|_2\left)\|(-\nabla^2)^{k+1/2}\Lamv^M\|_2
\phantom{\frac{\partial}{\partial}}\right.
\notag
\end{align}
implying
\begin{equation}
\|\Lamv\|_{2k+1,2}=\sup_M\|(-\nabla^2)^{k+1/2}\Lamv^M\|_2
\le\sum_{j=1}^3\left\|(-\nabla^2)^{k-1}\frac{\partial}{\partial x_j}
(\lamv\cdot\nabla)\Lamv\right\|_2+\|\fv\|_{2k-1,2}.
\label{ineq}\end{equation}

We need therefore to check that the norms in the sum at the right-hand side
of this inequality are bounded. By part (i) of the Theorem, the assumption
$\lamv\in W^{2k}_2(\Tbb^3)$ and $\Lamv\in W^{2k}_2(\Tbb^3)$ implies that $\lamv$
and $\Lamv$ and their derivatives of order up to $2k-2$ are continuous
(and hence uniformly bounded) vector fields in $\Tbb^3$. By the standard formula
for derivatives of products,
\begin{equation}
(-\nabla^2)^{k-1} \frac{\partial}{\partial x_j}(\lamv\cdot\nabla)\Lamv\end{equation}
is a linear combination of products of derivatives
\begin{equation}
\frac{\partial^{N_1}\lambda_q}{\partial^{n^1_1}x_1\partial^{n^1_2}x_2\partial^{n^1_3}x_3}
\,\frac{\partial^{N_2}\Lamv}{\partial^{n^2_1}x_1\partial^{n^2_2}x_2\partial^{n^2_3}x_3}\, ,
\label{prod}\end{equation}
where $N_i=n^i_1+n^i_2+n^i_3$, $0\le N_1\le 2k-1$, $N_2=2k-N_1$. Thus,
each of the terms is continuous and bounded, except maybe those for
$N_1=0,1$ or $2k-1$. If $N_1=0$, or $N_1=1$ or $2k-1$ for $k>1$,
one of the factors is continuous and the second one is known to belong to
$L_2(\Tbb^3)$, and hence their contributions to the right-hand side of \rf{ineq}
are finite. The remaining possibility is $k=N_1=N_2=1$, but in this case
both factors belong to $L_4(\Tbb^3)$ because $\lamv\in W^2_2(\Tbb^3)$ and
$\Lamv\in W^2_2(\Tbb^3)$ by the results of the previous subsection, and again
the respective norms are bounded.

We have thus established that $\Lamv\in W^{2k+1}_2(\Tbb^3)$.
By similar arguments from \rf{leqn} we find $\lamv\in W^{2k+1}_2(\Tbb^3)$.

To proceed, we scalar multiply in $L_2(\Tbb^3)$ \rf{Leqn} by
$(-\nabla^2)^{2k+2}\Lamv^M$, and obtain
\begin{align}
\int_{\Tbb^3}\Lamv\cdot(-\nabla^2)^{2k+2}\Lamv^M\dx
&+\int_{\Tbb^3}((-\nabla^2)^k(\lamv\cdot\nabla)\Lamv)\cdot(-\nabla^2)^{k+1}\Lamv^M\dx
\notag
\\
&=\int_{\Tbb^3}(-\nabla^2)^k\fv\cdot(-\nabla^2)^{k+1}\Lamv^M\dx,\end{align}
and therefore
\begin{equation}
\|\Lamv\|_{2k+2,2}=\sup_M\|(-\nabla^2)^{k+1}\Lamv^M\|_2
\le\|(-\nabla^2)^k(\lamv\cdot\nabla)\Lamv\|_2+\|\fv\|_{2k,2}.
\label{ineqtwo}\end{equation}
Since $\Lamv\in W^{2k+1}_2(\Tbb^3)$ and $\lamv\in W^{2k+1}_2(\Tbb^3)$,
by part (i) of the Theorem any derivative of $\Lamv$ and $\lamv$
of order up to $2k-1$ is continuous in $\Tbb^3$. Hence, in the expansion of
$(\lamv\cdot\nabla)\Lamv$ in a linear combination of products \rf{prod}
each term is either continuous, or a product of a continuous function by
a function from $L_2(\Tbb^3)$. Thus, \rf{ineqtwo} demonstrates that
$\Lamv\in W^{2k+2}_2(\Tbb^3)$. Similarly \rf{leqn} yields
$\lamv\in W^{2k+2}_2(\Tbb^3)$, concluding the demonstration.

Thus mathematical analysis of the problem yields both good and bad news.
The good news is that the problem \rf{eqLamv}, \rf{eqlamv} and \rf{lamsol}
necessarily has at least one classical solution, meaning in our case infinitely
differentiable at each point. Nothing is known about the number
of solutions except that it is strictly positive, nor is stability of any of the
MHD steady states guaranteed. The bad news is that the bounds
for the solutions and their derivatives rapidly degrade as $\eps\to 0$.
In particular, the inequalities that we have derived are insufficient
to claim that $\eps\nabla^2\lamv\to0$: the relevant $L_2$ bound we have derived is \rf{gap}.

It is interesting to compare this general result, that is for a general forcing,
with our numerical study of dissipative regions in the Archontis case.
For example, in \S\ref{ssecdiff} we find peak values
$\lamv=O(\eps^{-1/2})$ on scales of order $\eps^{1/2}$ indicating the scaling
$\eps\nabla^2\lamv=O(\eps^{-1/2})$. Note that these anomalously high
values are concentrated only in cylindrical cigars about the separatrices, of
radius $O(\eps^{1/2})$ and so occupy an $O(\eps)$ volume of space. This results
in the estimate $\|\lamv\|_{2,2}=O(\eps^{-1})$, which is a significantly milder
singularity than the one suggested by our bound \rf{gap}. The high values have
a negligible impact on the energy spectrum, there being no peak visible at small
scales. This can probably explain the gap between the `worst case scenario'
predicted by the rigorous mathematical analysis of the problem and
the numerical results: the Sobolev norms, that we have used, prove inefficient
in controlling formation of singularities in localised regions, because they
are of inherently integral nature. We should also note that our simulations
would not be able to resolve structure on scales much smaller than $O(\eps^{1/2})$.
%on the other hand, we do not see evidence of such structure emerging.

The apparent deterioration of the derivatives of the solution with their
order can be a spurious artefact due to imperfection of the proof
(which is especially possible in view of the generality of our arguments ---
at no point in \S\S \ref{maths}--\ref{smooth} we have made use of the fact
that the dynamo that we consider is powered by the Kolmogorov forcing
\rf{eqSvdef}), or a real
attribute of the solutions. It is likely that for some forcing in \rf{eqLamv}
and \rf{eqlamv} the worst case scenario suggested by these bounds is indeed
realised: they are based on the norm bounds provided by the embedding theorem,
which are sharp. In any case, this indicates that any naive approach to the
study of the limit $\eps\to0$ (for a general forcing) whereby the diffusive
term in \rf{eqlamv} is just discarded, is likely to be erroneous; this
can only be done in the region outside dissipative structures. In the absence
of the dominant elliptic operator, the equations obtained in this way are not in
general guaranteed to have solutions. When they exist, the solutions are likely
to develop singularities at some points or on certain manifolds, or possibly
on sets of a more complex structure. Note that locally the existence of solutions
is not a problem: the difficulty is in gluing together patches of such
solutions. The fast dynamo problem embodies a similar difficulty, with
small scales of magnetic field occurring in $O(1)$ volumes of space, though
with the fields concentrating on multifractal sets \citep{ChGi95}. Note
however, that the $L_2(\Tbb^3)$ norm of $\Lamv$ is uniformly (over $\eps$)
bounded, so the singularities are likely to be more pronounced
in the derivatives of the solution, rather than in the solution itself.

\section{Discussion}\label{secdisc}

We have presented investigations into the structure of the magnetic field and
flow in the equilibrated regime of the Archontis dynamo. Because of the highly
three-dimensional nature of the system, application of the available analytical
tools yields only rough results of limited value, and we
lack any kind of complete solution. What we have done is first to extend the
range of diffusivities $\eps$ over which the saturation mechanism operates to
give the steady state with nearly aligned fields. We have also classified the
symmetries of these flows and measured the field structure on the separatrices,
home of the cigar-like dissipative regions.

Then, using basic analytical tools, we have investigated the scaling of
diffusive terms near the separatrices. Here at leading order the field
$\lamv=\eps^{-1}\Lamv_-$ that enters from the body of the flow is transported along
characteristics of $\Lamv = \Lamv_+$. Where these characteristics come together
in the compressive flow at the stagnation points, where trajectories spiral in,
large gradients in $\lamv$ are generated, and diffusive terms enter the problem
on scales of $\eps^{1/2}$ as found by CG. In more general flows we may expect
a similar behaviour, with regions of heightened dissipation localised at points
where $\Lamv=0$ and along the unstable manifolds of such points. Of course in
the Archontis example the unstable manifolds link the stagnation points and so
the topology here is very simple and the dissipative regions very small, of
order $O(\eps)$ in volume: in other cases they may wander through the
three-dimensional space, giving a picture of much greater complexity, as could
be occurring in examples in \cite{CaGa06b}. Again wider regions of dissipation,
perhaps dense in the space, could occur if examples exist where $\Lamv$ has no
stagnation points; unfortunately the form of $\Lamv$ is not under our control
except where strongly constrained by symmetries. In order to cope with unknown
levels of geometrical complexity, an approach based on functional analysis is
appropriate, and this is the final part of the paper, in which the existence and
smoothness properties of steady solutions are established.

An analogy of the naively truncated equations (namely (\ref{eqLamv}, \ref{eqlamv}) with $\eps=0$) with the Euler equation, which is the subject of intense research, is instructive. A method for the investigation
of the evolutionary Euler and Navier--Stokes equations consists of the introduction
into the equations of new regularising terms, such as $\eps(-\nabla^2)^\alpha\uv$
or $\eps(-\nabla^2)^\beta(\partial\uv/\partial t)$. It has been known for decades
that for $\alpha>5/4$ solutions to the regularised equations are infinitely
differentiable at any $t>0$; for $\beta\geq1 /2$ and $\beta > 5/6$ one can prove
analyticity, at any $t>0$, of solutions to the Navier--Stokes and Euler
equations, respectively \citep{Zh10}. For any $\alpha$ or $\beta$ below the
respective thresholds, the problem is as difficult as the one for the original
equation. When the limit $\eps\to0$ is considered, the results are so far
inconclusive. One can only show that there exist sequences $\eps_k\to0$
such that solutions for these $\eps_k$ converge to a weak solution to the
non-regularised equation, and either the limit weak solution is unique for all
such sequences, or there exists a continuum of weak solutions. Whether for
$\eps\to0$ singularities develop in derivatives of the regularised solutions,
and how strong they are if they develop, remains unknown.

The difficulties arise in the general theory, because the bounds for solutions
are singular in $\eps$ as $\eps\to0$. Here the analogy with the Archontis
dynamo problem crystallises: in the Archontis problem the diffusive terms can
be regarded as a regularisation of the naively truncated diffusionless problem,
and we need to find out what happens when the regularisation parameter $\eps$
tends to zero. (In the diffusionless, i.e.\ non-regularised, case
it is unclear whether weak steady solutions exist.)
We note that the analogy may work both ways: the asymptotic analysis near
the separatrix in the Archontis dynamo (which we present in \S\S \ref{secflowfield} and \ref{secPDE})
may contain clues to what happens in solutions to the regularised Euler
(or even Navier--Stokes) equations in the limit $\eps\to0$. Unfortunately,
the clues are well hidden, because the regularising term in the Archontis
problem is of a different structure, and a very specific symmetric steady
solution to the general system of MHD equations is considered.

Besides further attempts to carry out an asymptotic analysis of
equations \rf{eqLamv} and \rf{eqlamv} and their evolutionary versions,
a number of other directions could be pursued in the future, for example
investigating time-dependent modifications to the steady Kolmogorov
forcing used here, or studying the evolution of superposed large-scale fields
and corresponding non-helical transport effects, as in the recent work of
\cite{SuBr09}.

\section*{Acknowledgements}

We are grateful for funding from the Royal Society/CNRS to enable
collaboration between ADG and YP, and for a Royal Society visitor
grant that supported the work of VZ at the University of Exeter.
ADG also thanks the Leverhulme Trust for their award of a Research
Fellowship, during which some of this research was undertaken. VZ was also
financed by the grants ANR-07-BLAN-0235 OTARIE from Agence Nationale de
la Recherche, France, and 07-01-92217-CNRSL{\_}a from the Russian foundation
for basic research. YP thanks A. Miniussi for computing design assistance.
Computer time was provided by GENCI (x2010021357) and
the Mesocentre SIGAMM machine, hosted by the Observatoire de la C\^ote d'Azur.

%The interest in this problem began at the
%Magnetohydrodynamics of Stellar Interiors programme at the Isaac Newton
%Institute, and we are grateful to the organisers for a valuable meeting.

We are grateful to Robert Cameron, David Galloway and Andrew Soward for
discussions and comments, particularly during the valuable Isaac Newton Institute (Cambridge, U.K.) programme on Magnetohydrodynamics of Stellar Interiors in 2004.
Finally we thank the authors of the open source computer package VAPOR; see {\tt www.vapor.ucar.edu}.


\begin{thebibliography}{}

\bibitem[Archontis(2000)]{Ar00}
Archontis, V. 2000
\emph{Study of generation and evolution of magnetic fields in stars using 3D MHD simulations of turbulent flows.}
PhD Thesis, Copenhagen University.

\bibitem[Archontis, Dorch \& Nordlund(2007)]{ArDoNo07}
Archontis, V., Dorch, S.B.F. \& Nordlund, {A}. 2007
Nonlinear MHD dynamo operating at equipartition.
\emph{Astron. \& Astrophys.} {\bf 472} 715--726.

\bibitem[Arnold \& Korkina(1983)]{ArKo83}
Arnold, V.I. \& Korkina, E.I. 1983
The growth of a magnetic field in a three-dimensional steady incompressible flow.
\emph{Vest. Mosk. Un. Ta. Ser. 1, Matem. Mekh.}, no.\ 3, 43--46.

%\bibitem[Beale, Kato \& Majda(1984)]{Majda}
%Beale, J.T., Kato, T. \& Majda, A. 1984
%Remarks on the breakdown of smoothness for the 3-D Euler equations.
%\emph{Comm.~Math.~Phys.} {\bf 94}, 61--66.

\bibitem[Bergh \& L\"ofstr\"om(1976)]{BL76}
Bergh, J. \& L\"ofstr\"om, J. 1976
\emph{Interpolation spaces}.
Springer-Verlag.

\bibitem[Cameron \& Galloway(2006a)]{CaGa06a}
Cameron, R. \& Galloway, D. 2006a
Saturation properties of the Archontis dynamo.
\emph{M. Not. R. Astr. Soc.} {\bf 365} 735--746.
Referred to as CG in the text.

\bibitem[Cameron \& Galloway(2006b)]{CaGa06b}
Cameron, R. \& Galloway, D. 2006b
High field strength modified ABC and rotor dynamos.
\emph{M. Not. R. Astr. Soc.} {\bf 367} 1163--1169.

\bibitem[Cattaneo, Hughes \& Kim(1996)]{CaHuKi96}
Cattaneo, F., Hughes, D.W. \& Kim, E.-J. 1996
Suppression of chaos in a simplified dynamo model.
\emph{Phys. Rev. Lett.} {\bf 76}, 2057--2060.

\bibitem[Childress \& Gilbert(1995)]{ChGi95}
Childress, S.  \&  Gilbert, A.D. 1995
\emph{Stretch, twist, fold: the fast dynamo}.
Springer-Verlag.

\bibitem[Childress, Kerswell \& Gilbert(2001)]{ChKeGi01}
Childress, S., Kerswell, R.R. \&  Gilbert, A.D. 2001
Bounds on dissipation for Navier--Stokes flow with Kolmogorov forcing.
\emph{Physica D}, {\bf 158}, 105--128.

\bibitem[Courvoisier, Hughes \& Proctor(2010)]{CoHuPr10}
Courvoisier, A., Hughes, D.W. \& Proctor, M.R.E 2010
Self-consistent mean-field magnetohydrodynamics
\emph{Proc. R. Soc. A} {\bf 466}, 583--601.

\bibitem[Dobrowolny, Mangeney \& Veltri(1980)]{DoMaVe80}
Dobrowolny, M., Mangeney, A. \& Veltri, P. 1980
Fully developed anisotropic hydromagnetic turbulence in interplanetary space.
\emph{Phys. Rev. Lett.} {\bf 45}, 144--147.

\bibitem[Dorch \& Archontis(2004)]{DoAr04}
Dorch, S.B.F. \& Archontis, V. 2004
On the saturation of astrophysical dynamos:
numerical experiments with the no-cosines flow.
\emph{Solar Phys.} {\bf 224} 171--178.
Referred to as DA in the text.

\bibitem[Dombre \emph{et al.}(1986)]{DoFrGrHeMeSo86}
Dombre, T. Frisch, U., Greene, J.M., H\'enon, M., Mehr, A. \& Soward, A.M. 1986
Chaotic streamlines in the ABC flows.
\emph{J. Fluid Mech.} {\bf 167}, 353--391.

\bibitem[Friedlander \& Vishik(1995)]{FrVi95}
Friedlander, S. \& Vishik. M.M. 1995
On stability and instability criteria for magnetohydrodynamics
\emph{Chaos} {\bf 5}, 416--423.

\bibitem[Galloway \& Proctor(1992)]{GaPr92}
Galloway, D.J. \& Proctor, M.R.E. 1992
Numerical calculations of fast dynamos in smooth velocity fields with realistic diffusion.
\emph{Nature} {\bf 356}, 691--693.

\bibitem[Gilbert(1992)]{Gi92}
Gilbert, A.D. 1992
Magnetic field evolution in steady chaotic flows.
\emph{Phil. Trans. R. Soc. Lond. A} {\bf 339}, 627--656.

\bibitem[Ladyzhenskaya(1969)]{Lad69}
Ladyzhenskaya, O.A. 1969
\emph{The mathematical theory of viscous incompressible flow}.
Gordon and Breach, New York, London.

\bibitem[Leray \& Schauder(1934)]{LeSch}
Leray, J. \& Schauder, J. 1934
Topologie et equations fonctionelles.
\emph{Ann. Sci. \'Ecole Norm. Sup.} {\bf 13}, 45--78.

\bibitem[Mason, Cattaneo \& Boldyrev(2006)]{MaCaBo06}
Mason, J., Cattaneo, F. \& Boldyrev, S. 2006
Dynamic alignment in driven magnetohydrodynamic turbulence
\emph{Phys. Rev. Lett.} {\bf 97}, 255002.

\bibitem[Pouquet, Meneguzzi \& Frisch(1986)]{PoMeFr86}
Pouquet, A., Meneguzzi, M. \& Frisch, U. 1986
Growth of correlations in magnetohydrodynamic turbulence
\emph{Phys. Rev. A} {\bf 33}, 4266--4276.

\bibitem[Sur \& Brandenburg(2009)]{SuBr09}
Sur, S. \& Brandenburg, A. 2009
The role of the Yoshizawa effect in the Archontis dynamo.
\emph{Mon. Not. R. Astron. Soc.} {\bf 399}, 273--280.

\bibitem[Taylor(1981)]{Tay81}
Taylor, M. 1981
\emph{Pseudodifferential operators}.
Princeton University Press.

\bibitem[Zheligovsky(2009)]{Zh09}
Zheligovsky, V. 2009
Determination of a flow generating a neutral magnetic mode.
\emph{Phys. Rev. E}, {\bf 80}, 036310.

\bibitem[Zheligovsky(2010)]{Zh10}
Zheligovsky, V. 2010 A priori bounds for Gevrey--Sobolev norms of
space-periodic three-dimensional solutions to equations of hydrodynamic type.
\emph{Differential and integral equations}, submitted
(arXiv:1001.4237 [math.AP]).

\bibitem[Zienicke, Politano \& Pouquet(1998)]{ZiPoPo98}
Zienicke, E., Politano, H.  \& Pouquet, A. 1998
Variable intensity of Lagrangian chaos in the nonlinear dynamo problem.
\emph{Phys. Rev. Lett.} {\bf 81}, 4640--4643.

\end{thebibliography}
\end{document}